\documentclass[aps,prd,nofootinbib,preprint]{revtex4}
\usepackage{hyperref}
\usepackage{ulem}
\usepackage{color}
\usepackage{graphicx}
\usepackage{multirow}
\usepackage{amsmath,amssymb,amsthm,amsxtra,overpic,bbm,bm,float
,epsfig}

\usepackage{color}

\fontsize{3}{5}\selectfont

\begin{document}
\title{Study of a tri-direct littlest seesaw model at MOMENT}
%\title{A case study of flavour symmetry at MOMENT: the tri-direct littlest seesaw model}
%%%%%%%%%%%%%%%%%%%%%%%%%%%%%%%%%%%%%%%%%%%%%%%%%%%%
\author{Jian Tang$^1$\footnote{tangjian5@mail.sysu.edu.cn}}
\author{Tse-Chun Wang$^1$\footnote{wangzejun@mail.sysu.edu.cn}}
\affiliation{$^1$School of Physics, Sun Yat-Sen University, Guangzhou 510275, China}

\begin{abstract}
The flavour symmetry succeeds in explaining the current global fit results. Flavour-symmetry models can be tested by the future experiments that improve the precision of neutrino oscillation parameters, \textit{such as} the MuOn-decay MEdium baseline NeuTrino beam experiment (MOMENT). In this work, we consider tri-direct littlest seesaw (TDLS) models for a case study, and analyze how much MOMENT can extend our knowledge on the TDLS model. We find that measurements of $\theta_{23}$ and $\delta$ are crucial for MOMENT to exclude the model at more than $5\sigma$ confidence level, if the best fit values in the last global analysis result is confirmed. Moreover, the $3\sigma$ precision of model parameters can be improved at MOMENT by at least a factor of two. Finally, we project the surface at the $3\sigma$ confidence level from the model-parameter space to the oscillation-parameter space, and find the potential of MOMENT to observe the sum rule between $\theta_{23}$ and $\delta$ predicted by TDLS.
\end{abstract}
%\pacs{13.15.+g, 14.60.Pq, 14.60.St}

\maketitle
%\tableofcontents
\section{Introduction}\label{intro}

The discovery of neutrino oscillations points out the fact that neutrinos have mass, and provides evidence beyond the Standard Model (BSM). This phenomenon is successfully described by a theoretical framework with the help of three neutrino mixing angles ($\theta_{12}$, $\theta_{13}$, $\theta_{23}$), two mass-square splittings ($\Delta m_{21}^2$, $\Delta m_{31}^2$), and one Dirac CP phase ($\delta$) \cite{Pontecorvo:1967fh,Maki:1962mu,Pontecorvo:1957qd,Esteban:2018azc}. Thanks to the great efforts in the past two decades, we almost have a complete understanding of such a neutrino oscillation framework. More data in the neutrino oscillation experiments is needed to determine the sign of $\Delta m_{31}^2$, to measure the value of $\sin\theta_{23}$, to discover the potential CP violation in the leptonic sector and even to constrain the size of $\delta$ \cite{Esteban:2018azc}. For these purposes, the on-going long baseline experiments (LBLs), such as the NuMI Off-axis $\nu_e$ Appearance experiment (NO$\nu$A)~\cite{Ayres:2007tu} and the Tokai-to-Kamioka experiment (T2K)~\cite{Abe:2011ks}, can answer these questions with the statistical significance $\gtrsim 3\sigma$ in most of the parameter space. Based on the analysis with their data, the normal mass ordering ($\Delta m_{31}^2>0$), the higher $\theta_{23}$ octant ($\theta_{23}>45^\circ$), and $\delta\sim270^\circ$ are preferred so far~\cite{Esteban:2018azc}. The future LBLs, Deep Underground Neutrino Experiment (DUNE)~\cite{Acciarri:2015uup}, Tokai to Hyper-Kamiokande (T2HK)~\cite{Abe:2014oxa}, and the medium baseline reactor experiment, the Jiangmen Underground Neutrino Observatory (JUNO)~\cite{Djurcic:2015vqa,An:2015jdp} will further complete our knowledge of neutrino oscillations.

The MuOn-decay MEdium baseline NeuTrino beam experiment (MOMENT) has been proposed and is under consideration. Apart from superbeam neutrino experiments like DUNE or T2HK, it is planned to be at muon-decay accelerator neutrino experiments. In such experiments, neutrinos come from a three-body decay process, avoiding intrinsic electron-flavor neutrino contaminations in the reconstructed signals from the source. In addition, MOMENT~\cite{Cao:2014bea} is likely to use a Gd-doped water Cherenkov detector, which is capable of detecting multiple channels. MOMENT is understood to have excellent properties to study BSM physics, \textit{e.g.} the invisible $\nu_3$ decay~\cite{Tang:2018rer}, NSIs~\cite{Gavela:2008ra,Bonnet:2009ej,Krauss:2011ur} and sterile neutrinos~\cite{Gariazzo:2017fdh,Abazajian:2012ys,Adhikari:2016bei,Minkowski:1977sc}. Though the current studies on MOMENT have mainly focused on other BSM physics~\cite{Tang:2017qen,Tang:2017khg}, it is also necessary to perform physics study related to the standard neutrino oscillation to test the flavour symmetry models.

The symmetry of discrete groups, preserved at the high energy but slightly broken at the lower energy, predicts the neutrino mixing, mass-square splittings, and the CP violation phase (Dirac and Majorana phases), with reduced degrees of freedom (some of useful review articles are~\cite{Altarelli:2010gt,Ishimori:2010au,King:2013eh,King:2014nza,King:2015aea,King:2015ata,King:2017guk}). As a result, these models do not only simplify the theoretical framework for neutrino oscillations, but also provide a theoretical reason for this phenomenon. Many of these models can well describe the current neutrino-oscillation data. One of the most predictive models is the littlest seesaw model (LSS), which includes two massive right-handed neutrinos: one corresponds to the atmospheric-mass term, while the other is included for the solar-mass term~\cite{King:2013iva,King:2015dvf,King:2016yvg}.  
The littlest seesaw model in the tri-direct approach (TDLS) has been proposed and succeeds in describing the current global-fit results~\cite{Ding:2018fyz,Ding:2018tuj}. In this model, four parameters $x$, $\eta$, $r$, $m_a$ are used to describe neutrino oscillations. This model has been studied with simulated data at NO$\nu$A, T2K, DUNE, T2HK and JUNO~\cite{Ding:2019zhn}. In this work, we study how the next-generation neutrino project using muon-decay beams such as MOMENT can further extend our knowledge on the TDLS model. 

This paper is arranged as follows. In Sec.~\ref{sec:model}, we will introduce how TDLS models predict oscillation parameters, before presenting how this model describes the NuFit4.0 result. In Sec.~\ref{sec:simulation}, we will introduce the statistics and simulation details used in this paper. We will show the definition of $\chi^2$, including the way that we implement ``the pull method'' to estimate the impact of systematic uncertainties, and how we include the current global-fit results by priors. We will then summarize the assumed configurations for the MOMENT experiment, and will show how the probabilities for MOMENT will be changed by varying each of model parameters. The simulation results will be shown in Sec.~\ref{sec:results}. We will present the model exclusion capability at MOMENT and how model parameters can be constrained by MOMENT data. We will discuss results of projecting the $3\sigma$ sphere from the model-parameter space to the standard-parameter space. Finally, we will close up this paper in Sec.\ref{sec:conclusion} with our conclusions.

\section{Model review: littlest Seesaw in the Tri-Direct approach}\label{sec:model}

% \usepackage{multirow}
\begin{table}[h!]
\caption{\label{tab:parameters}A summary of the relation between oscillation parameters and TDLS model parameters~\cite{Ding:2018fyz}. Two requirements are imposed by TDLS: the smallest mass state $m_1=0$ and the normal mass ordering. The sign of $\sin\delta$ depends on the sign of $x\cos\psi$: ``$+$'' (``$-$'') is for $x\cos\psi>0$ ($<0$).}
\begin{tabular}{l|l}
\hline\hline
model parameters                                 & $x$, $\eta$, $r$, $m_a$                                                                                                                                                                         \\\hline
\multirow{6}{*}{combinations of model parameters} & $y=\frac{5x^2+2x+2}{2\left(x^2+x+1\right)}(m_{a}+e^{i \eta } m_{s})$,                                                                                                                            \\
                                                 & $z=-\frac{\sqrt{5x^2+2x+2}}{2\left(x^2+x+1\right)}\left[ (x+2)m_{a}-x(2x+1)e^{i \eta }m_{s}\right]$,                                                                                             \\
                                                 & $w=\frac{1}{2(x^2+x+1)}\left[(x+2)^2m_{a}+x^2\left(2x+1\right)^2e^{i \eta } m_{s}\right]$,                                                                                                       \\
                                                 & $\sin\psi=\frac{\Im\left(y^{*}z+wz^{*}\right)}{|y^{*}z+wz^{*}|},\quad \cos\psi=\frac{\Re\left(y^{*}z+wz^{*}\right)}{|y^{*}z+wz^{*}|}$.                                                           \\
                                                 & \begin{tabular}[c]{@{}l@{}}$\sin2\theta=\frac{2|y^{*}z+wz^{*}|} {\sqrt{(|w|^2-|y|^2)^2+4|y^{*}z+wz^{*}|^2}},$\end{tabular}                                                                     \\
                                                 & $\cos2\theta=\frac{|w|^2-|y|^2}{\sqrt{(|w|^2-|y|^2)^2+4|y^{*}z+wz^{*}|^2}}$.                                                                                                                     \\\hline
\multirow{7}{*}{oscillation parameters}          & $\Delta m_{21}^2=m^2_2=\frac{1}{2}\left[\left|y\right|^2+\left|w\right|^2+2\left|z\right|^2-\frac{\left|w\right|^2-\left|y\right|^2}{\cos\theta}\right]$,                                        \\
                                                 & $\Delta m_{31}^2=m^2_3=\frac{1}{2}\left[\left|y\right|^2+\left|w\right|^2+2\left|z\right|^2+\frac{\left|w\right|^2-\left|y\right|^2}{\cos\theta}\right]$,                                        \\
                                                 & $\sin^2\theta_{12}=1-\frac{3x^2 }{3x^2+2\left(x^2+x+1\right) \cos^2\theta }$,                                                                                                                  \\
                                                                                                  & $\sin^2\theta_{13}=\frac{2\left(x^2+x+1\right)\sin^2\theta}{5x^2+2x+2}$,                                                                                                                       \\
                                                 & $\sin^2\theta_{23}=\frac{1}{2}+\frac{x\sqrt{3\left(5x^2+2x+2\right)}\sin2\theta\sin\psi }{2\left[3x^2+2\left(x^2+x+1\right) \cos^ 2 \theta\right]}$,                                          \\
                                                 & $\cos\delta=\frac{ \cot 2 \theta_{23} \left[3x^2-\left(4x^2+ x+1\right)\cos^2\theta_{13}\right]}{\sqrt{3} \left|x\right| \sin \theta_{13} \sqrt{\left(5x^2+2x+2\right)\cos^2\theta_{13}-3x^2}}$, \\
                                                 & $\sin\delta= \pm\csc 2 \theta_{23} \sqrt{1+\frac{\left(x^2+x+1\right)^2 \cot ^2\theta_{13} \cos ^22 \theta_{23}}{3x^2 \left[3x^2 \tan ^2\theta_{13}-2 \left(x^2+x+1\right)\right]}}$.   \\\hline \hline       
\end{tabular}
\end{table}

The littlest seesaw model in the tri-direct approach is currently proposed, and succeeds in describing the current neutrino-oscillation data~\cite{Ding:2018fyz}. In this model, the atmospheric and solar flavon vacuum alignments are $\langle\phi_{\text{atm}}\rangle\propto\left(1, \omega^2, \omega\right)^T$ and $\langle\phi_{\text{sol}}\rangle\propto\left(1, x, x\right)^T$,
where $\omega=e^{2\pi i/3}$ stands for a cube root of unity and the parameter $x$ is real because of the imposed CP symmetry. As a result, the Dirac neutrino mass matrix reads as follows:
\begin{equation}
m_{D}=\begin{pmatrix}
y_{a}      ~&~    y_{s}  \\
\omega y_{a}  ~&~  x y_{s} \\
\omega^2y_{a}  ~&~  x y_{s}
\end{pmatrix} \,.
\end{equation}
The right-handed neutrino Majorana mass matrix is diagonal
\begin{equation}
m_{N}=\begin{pmatrix}
M_{\textrm{atm}}  &  0  \\
0  &   M_{\textrm{sol}}
\end{pmatrix}\,.
\end{equation}
%The light effective Majorana neutrino masses are generated via the seesaw mechanism such that the light left-handed Majorana neutrino mass matrix is given by
Under the littlest seesaw model, the light left-handed Majorana neutrino mass matrix is given by
\begin{equation}
\label{eq:mnu}  m_{\nu}=m_{a}\begin{pmatrix}
 1 &~ \omega  &~ \omega ^2 \\
 \omega  &~ \omega ^2 &~ 1 \\
 \omega ^2 &~ 1 &~ \omega  \\
\end{pmatrix}+e^{i\eta}m_{s}
\begin{pmatrix}
 1 &~  x &~  x \\
 x &~ x^2 &~ x^2 \\
 x &~ x^2 &~ x^2 \\
\end{pmatrix}\,,
\end{equation}
where $m_a=|y^2_a/M_{\text{atm}}|$, $m_s=|y^2_s/M_{\text{sol}}|$, and the only physically important phase $\eta$ depends on the relative phase between $y^2_a/M_{\text{atm}}$ and $y^2_s/M_{\text{sol}}$. Obviously, from Eq.~(\ref{eq:mnu}), $m_1=0$ and the normal mass ordering are imposed by TDLS. We summarise the dependence of oscillation parameters on model parameters in Table~\ref{tab:parameters}. %From the expansions of $\sin^2\theta_{12}$ and $\sin^2\theta_{13}$ in Tab.~\ref{tab:parameters}, an interesting sum rule can be found,
Ref.~\cite{Ding:2018fyz} further predicts the sum rule for TDLS,
\begin{equation}\label{eq:correlation_mix_angles}
\cos^2\theta_{12}\cos^2\theta_{13}=\frac{3x^2}{5 x^2+2x+2}\,.
\end{equation}

\begin{table}[!h]
\caption{The best fit and $3\sigma$ uncertainty, in the results of NuFit4.0 \cite{Esteban:2018azc}.}
\begin{tabular}{|c|c|c|c|c|c|c|}
\hline
Parameter & $\theta_{12}/^\circ$ & $\theta_{13}/^\circ$ & $\theta_{23}/^\circ$ & $\delta/^\circ$  & $\Delta m_{21}^2/10^{-5}\text{eV}^2$ & $\Delta m_{31}^2/10^{-3}\text{eV}^2$\\ \hline
best fit & $33.82$ & $8.61$ & $49.6$ & $215$ & $7.39$ & $2.525$ \\\hline
$3\sigma$ Range & $31.61-36.27$ & $8.22-8.99$ & $40.3-52.4$ & $125-392$ & $6.79-8.01$ & $2.47-2.625$ \\\hline
\end{tabular}
\label{tab:nufit4.0}
\end{table}

\begin{table}[!h]
%\scriptsize{
\caption{The best fit for $x$, $\eta$, $r$, $m_a$ with the result of NuFit4.0 \cite{Esteban:2018azc}, and the corresponding oscillation parameters.}\label{tab:model_fit}
\begin{tabular}{|c||c|c|c|c||c|c|c|c|c|c|}
\hline
$\Delta\chi^2$ &$x$ & $\eta/\pi$ & $r$ & $m_a/$ meV & $\theta_{12}/^\circ$ & $\theta_{13}/^\circ$ & $\theta_{23}/^\circ$ & $\delta/^\circ$  & $\Delta m_{21}^2/10^{-5}\text{eV}^2$ & $\Delta m_{31}^2/10^{-3}\text{eV}^2$\\ \hline
$4.98$ & $ -3.65$ & $1.13$ & $0.511$ & $3.71$ & $35.25$ & $8.63$ & $46.98$ & $278.96$ & $7.39$ & $2.525$\\ \hline
\end{tabular}
%}
\end{table}

We use the best fit value and the $3\sigma$ uncertainty of NuFit4.0 \cite{Esteban:2018azc} (shown in Table~\ref{tab:nufit4.0}), we find the best fit results for TDLS models in Table~\ref{tab:model_fit}. The $3\sigma$ uncertainty is given as
 \begin{equation}\label{eq:3sigma_nufit4}
 \begin{array}{c}
 -5.475<x<-3.37,~0.455<\eta/\pi<1.545,\\
 ~0.204<r<0.606,~3.343<m_a/\text{meV}<4.597.
 \end{array}
 \end{equation}
Notable between Tables~\ref{tab:nufit4.0} and \ref{tab:model_fit} is that the most inconsistent oscillation parameters are $\theta_{23}$ and $\delta$. The others are placed within the $1\sigma$ error, or even at the best-fit value (\textit{e.g.} $\Delta m_{21}^2$ and $\Delta m_{31}^2$). As a result, we are looking forward to improving precision measurements on $\theta_{23}$ and $\delta$ for further understanding of this model.

\section{Simulation details}\label{sec:simulation}

\subsection{Statistics Method}\label{sec:statistics}

\begin{figure}[!h]%
\centering
\includegraphics[width=5.5in]{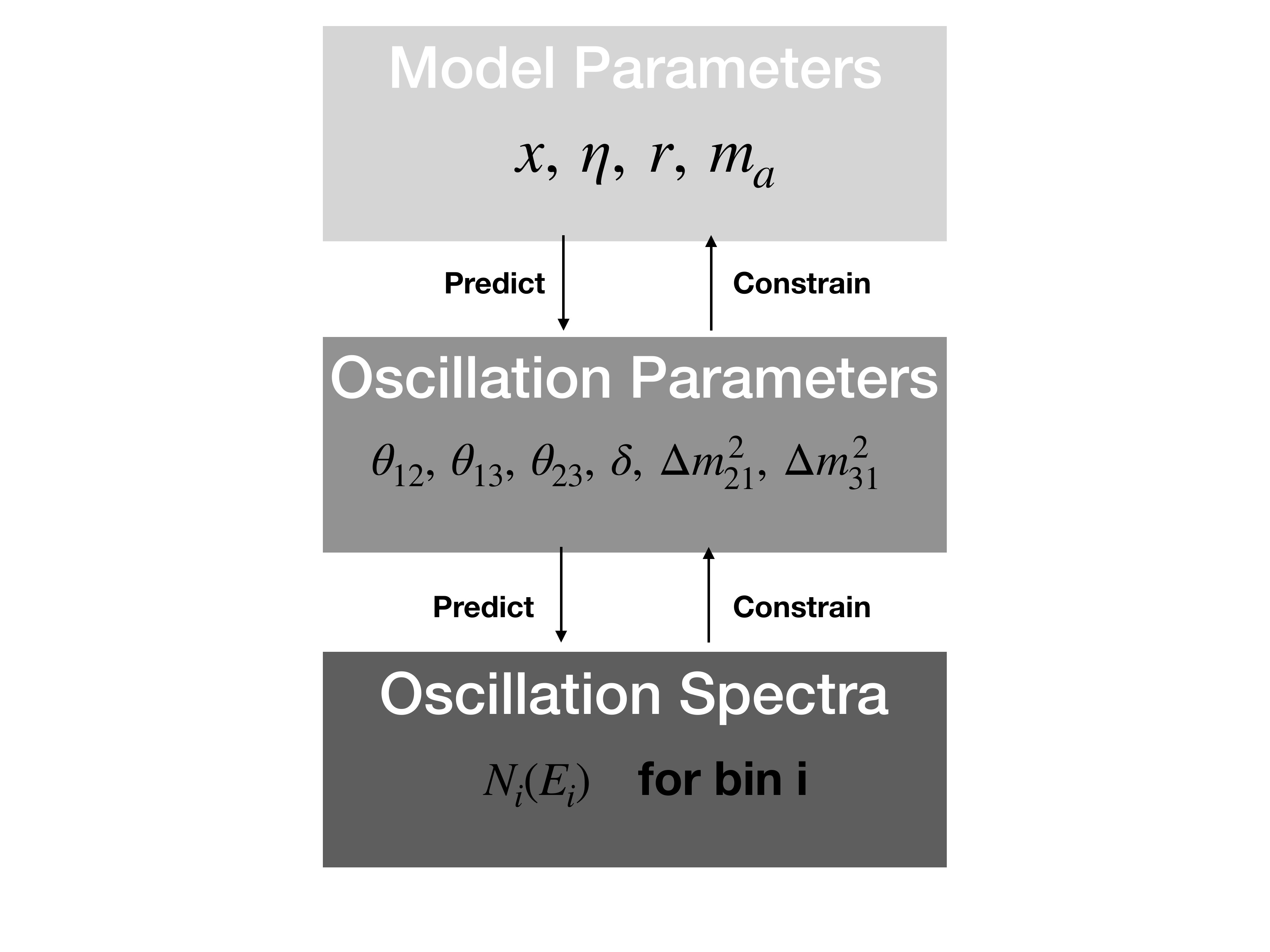}
\caption{A scheme to correlate the model parameters with standard neutrino oscillation parameters. The error propagation is implemented in the simulation code up to the spectra analysis.}%
\label{fig:parameters}
\end{figure}

The statistical study on the TDLS model at MOMENT can be understood in Fig.~\ref{fig:parameters}. The model imposes correlations between or among the standard neutrino oscillation parameters, and predicts the oscillation spectra for MOMENT. In other words, the neutrino spectra of MOMENT can constrain the standard oscillation parameters, and therefore test the TDLS model or constrain the model parameters. Based on this perspective, we use two methods to conduct the numerical analysis with the simulated data:
\begin{itemize}
 \item {The standard three neutrino oscillations expressed by three mixing angles, two mass-square splittings and one Dirac-CP phase: $\overrightarrow{\mathcal{O}}=\{\theta_{12}\,,\theta_{13}\,,\theta_{23}\,,\delta_\mathrm{CP}\,,\Delta m_{21}^2\,, \Delta m_{31}^2 \}$. We expect that precision measurements of mixing parameters are correlated with uncertainties of current global fit results. 
%It then predicts the expected number of events in each bin $\mu_i(\overrightarrow{\mathcal{O}})$. 
We suppose that a given experiment reconstructs neutrino spectra in $N$ bins sequentially. The number of observed events in the bin $i$ is recorded as $n_i$, which in our work is predicted by the true model. We can build a $\chi^2_{st.}(\overrightarrow{\mathcal{O}})$ to quantify the sensitivity:
\begin{equation}\label{eq:chi_PMNS}
 \chi^2_{st.}(\overrightarrow{\mathcal{O}})=\sum_{i=1}^N \left[ \frac{\mu_i(\overrightarrow{\mathcal{O}})-n_i}{\sigma_i} \right]^2\,,
\end{equation}
where $\mu_i$ is the number rate of bin $i$ predicted by the hypothesis $\overrightarrow{\mathcal{O}}$.
 }
 \item {We consider the following parameters from TDLS: $\overrightarrow{\mathcal{M}}=\{x\,,\eta\,,m_{a}\,,r \}$. 
Other steps in the likelihood analysis will follow the same strategy as the above method, but replace the equation Eq.~(\ref{eq:chi_PMNS}) with
\begin{equation}
\chi^2_{st.}(\overrightarrow{\mathcal{M}})=\sum_{i=1}^N \left[ \frac{\mu_i(\overrightarrow{\mathcal{O}}(\overrightarrow{\mathcal{M}}))-n_i}{\sigma_i} \right]^2\,,
\end{equation}
with standard oscillation parameters as a function of model parameters $\overrightarrow{\mathcal{O}}(\overrightarrow{\mathcal{M}})$.
%We can expect better measurements of input parameters after a combination of experimental results and symmetry-induced constraints from the theory.
}
\end{itemize}

To describe the impact of systematic uncertainties, we adopt the following modification:
\begin{equation}\label{eq:chi_sys}
\chi_{sys.}^2(\overrightarrow{\mathcal{O}}~\text{or}~\overrightarrow{\mathcal{M}})=\min_{\{\xi_s,\xi_b\}}
\sum_{i=1}^N \left[ \frac{\mu_i((\overrightarrow{\mathcal{O}}~\text{or}~\overrightarrow{\mathcal{O}}(\overrightarrow{\mathcal{M}});~\xi_s,\xi_b)-n_i}{\sigma_i} \right]^2+p(\xi_s,\sigma_s)+p(\xi_b,\sigma_b)\,.
\end{equation}
where $p(\xi,\sigma)=\xi^2/\sigma^2$ is a Gaussian prior on the nuisance parameter $\xi$ with the uncertainty $\sigma$ (subscripts $s$ and $b$ denote signal and background respectively)
and $\mu_i((\overrightarrow{\mathcal{O}}~\text{or}~\overrightarrow{\mathcal{O}}(\overrightarrow{\mathcal{M}});~\xi_s,\xi_b)$ is predicted event rate for bin $i$
\begin{equation}
\mu_i((\overrightarrow{\mathcal{O}}~\text{or}~\overrightarrow{\mathcal{O}}(\overrightarrow{\mathcal{M}});~\xi_s,\xi_b)=(1+\xi_s)\times\mu_{s,i}+(1+\xi_b)\times\mu_{b,i},
\end{equation}
with the signal rate $\mu_{s,i}$ and the background rate $\mu_{b,i}$ for each energy bin $i$.

To include the currently constraints for the neutrino oscillation parameters, we finally use 
\begin{equation}\label{eq:chi_TD}
\chi^2(\overrightarrow{\mathcal{O}}~\text{or}~\overrightarrow{\mathcal{M}})=\min_{\overrightarrow{\mathcal{O}}~\text{or}~\overrightarrow{\mathcal{M}}}\chi_{sys.}^2(\overrightarrow{\mathcal{O}}~\text{or}~\overrightarrow{\mathcal{M}})+\sum_i p(\overrightarrow{\mathcal{O}}_i(\overrightarrow{\mathcal{M}}),\overrightarrow{\mathcal{O}}_{cen.,i},\overrightarrow{\sigma}_i)\,,
\end{equation}
where $\sum_i p(\overrightarrow{\mathcal{O}}_{hyp.},\overrightarrow{\mathcal{O}}_{cen.},\overrightarrow{\sigma})$ is the summation of Gaussian priors over all oscillation parameters with two vectors: one includes all central values $\overrightarrow{\mathcal{O}}_{cen.}$ and the other consists of the standard deviation $\overrightarrow{\sigma}$. The values for $\overrightarrow{\mathcal{O}}_{cen.}$ and $\overrightarrow{\sigma}$ are taken from the best-fit value and according to $3\sigma$ uncertainties of the NuFit4.0 result~\cite{Esteban:2018azc} (shown in Table~\ref{tab:nufit4.0}), respectively. In this work, the values of $\overrightarrow{\mathcal{O}}(\overrightarrow{\mathcal{M}})$ are predicted by the TDLS model.

\subsection{Experiment Setting}
\begin{table}[!h]
\caption{Assumptions for the source, detector and the running time at MOMENT in the simulation.}
{\begin{tabular}{@{}ccc@{}}
\toprule
\multicolumn{2}{c}{MOMENT}\\
%Experiments & MOMENT \\
\colrule
Fiducial mass\hphantom{00} & \hphantom{0}Gd-doping Water cherenkov(500 kton) \\
\colrule
Channels\hphantom{00} & \hphantom{0}{$\nu_e(\bar{\nu}_e)\rightarrow\nu_e(\bar{\nu}_e)$, $\nu_{\mu}(\bar{\nu}_{\mu})\rightarrow\nu_{\mu}(\bar{\nu}_{\mu})$,} \\

&$\nu_e(\bar{\nu}_e)\rightarrow\nu_{\mu}(\bar{\nu}_{\mu})$, 
$\nu_{\mu}(\bar{\nu}_{\mu})\rightarrow\nu_e(\bar{\nu}_{e})$\\
\colrule
Energy resolution\hphantom{0} & $12\%/E$ \\
\colrule
Runtime & $\mu^-$ mode 5 yrs+ $\mu^+$ mode 5 yrs \\
\colrule
Baseline & 150 km \\
\colrule
Energy range  & 100 MeV to 800 MeV \\
\colrule
Normalization & appearance channels: $2.5\%$ \\
(error on signal) & disappearance channels: 5$\%$\\
\colrule
Sources of & {Neutral current, Atmospheric neutrinos}\\
Background& Charge misidentification\\
\botrule
\end{tabular}
}
\label{tab:glbtable}
\end{table}

We summarize the simulation details for MOMENT in Table~\ref{tab:glbtable}. MOMENT, as a medium muon decay accelerator neutrino experiment, has been originally proposed as a future experiment to measure the leptonic CP-violating phase, though it also has good sensitivities on $\theta_{13}$, $\theta_{23}$ and $\Delta m^2_{31}$~\cite{Tang:2019wsv}.

The neutrino fluxes are kindly provided by the MOMENT working group~\cite{Cao:2014bea}. The events are taken from $100$ to $800$ MeV. We assume five-year data taken at the $\mu^-$ and $\mu^+$ mode, respectively. Eight oscillation channels ($\nu_e\rightarrow \nu_e$, $\nu_e\rightarrow \nu_{\mu}$, $\nu_{\mu} \rightarrow \nu_e$, $\nu_{\mu} \rightarrow \nu_{\mu}$ and their CP-conjugate partners) are considered in this work. Multi-channel analyses are helpful in measuring the values of multiple parameters. As a result, the detector design is also crucial to precisely read out the  events from different neutrino-oscillation channels. We have to consider flavour and charge identifications to distinguish secondary particles by means of an advanced neutrino detector --- a 500 kton Gd-doped water cherenkov detector. The charged-current interactions are used to identify neutrino signals: $\nu_e + n \rightarrow p + e^-$, $\bar{\nu}_{\mu} + p \rightarrow n + \mu^+$, $\bar{\nu}_e + p \rightarrow n + e^+$, and $\nu_{\mu} + n \rightarrow p + \mu^- $, with the new technology using Gd-doped water to separate both Cherenkov and coincident signals from capture of thermal neutrons~\cite{Campagne:2006yx,Ishida:2013kba}. The energy resolution is assumed $12\%/E$ for all channels. For the systematic uncertainties, we assume $\sigma_s=2.5\%$ for signal normalizations and $\sigma_b=5\%$ for background fluctuations.

The major background components come from the atmospheric neutrinos, neutral current backgrounds and charge mis-identifications. They can be largely suppressed with the beam direction and a proper modelling background spectra during the beam-off period, which are to be extensively studied in detector simulations. 
We consider matter effects during neutrino propagations with the help of the Preliminary Reference Earth Model (PREM) density profile is considered in the numerical calculations~\cite{prem:1981}.

\subsection{Neutrino oscillation probabilities in the TDLS model}

\begin{figure}[!h]
 \centering
\includegraphics[width=0.45\textwidth]{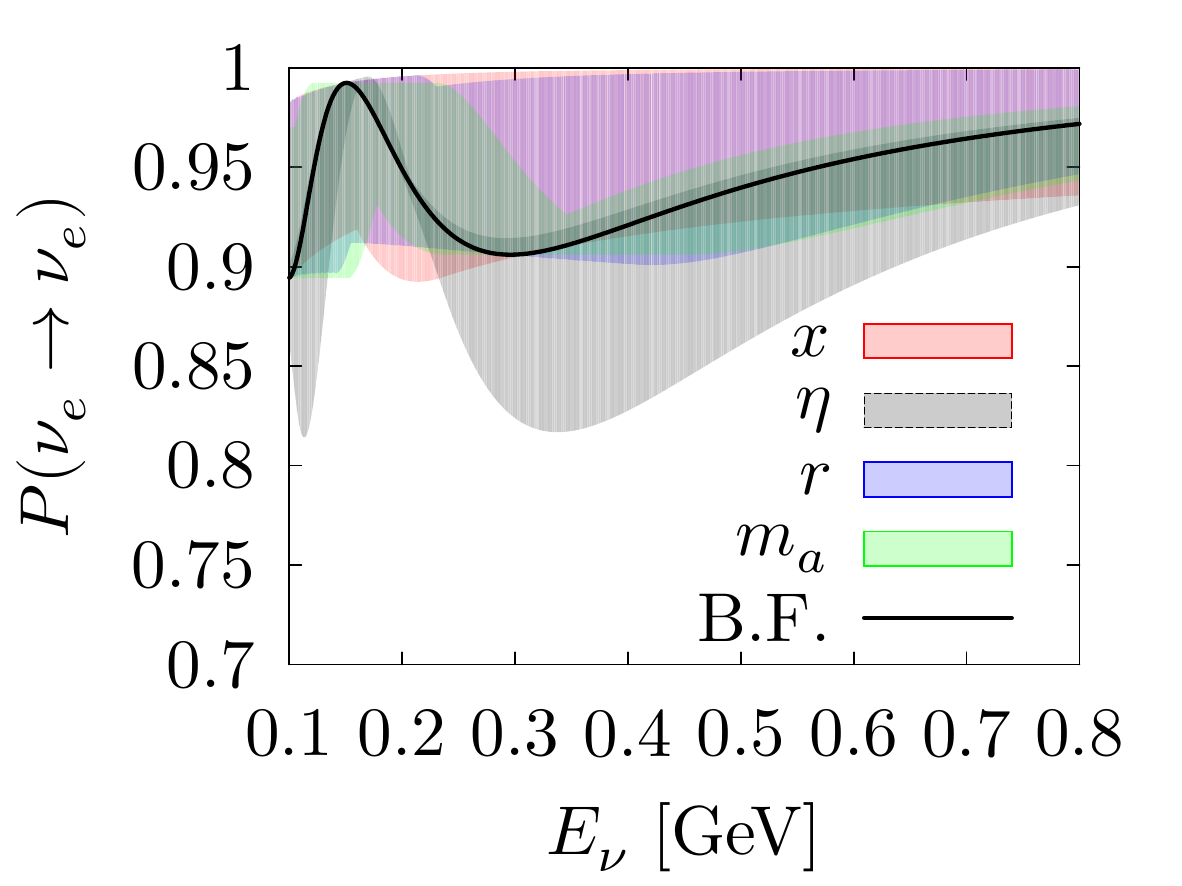}
 \includegraphics[width=0.45\textwidth]{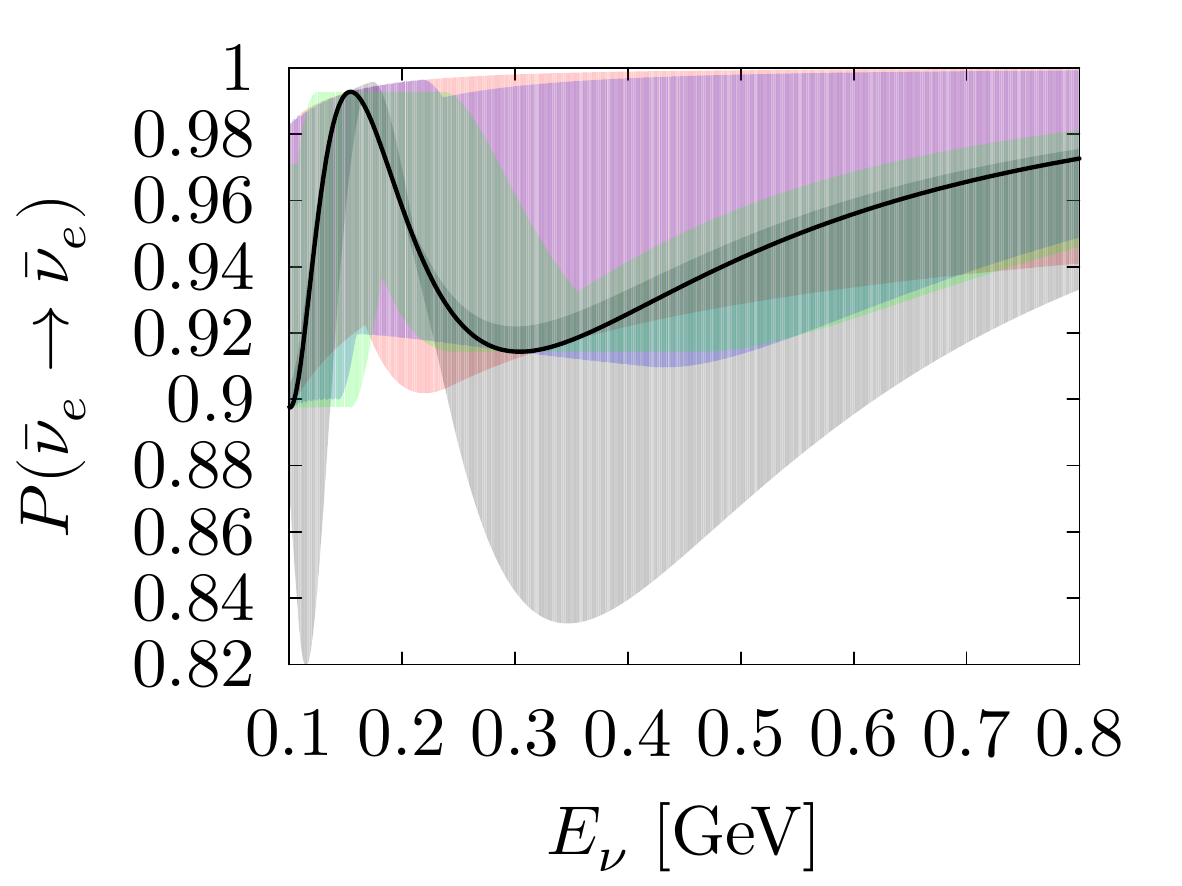}$~~~~~~$\\
 \includegraphics[width=0.45\textwidth]{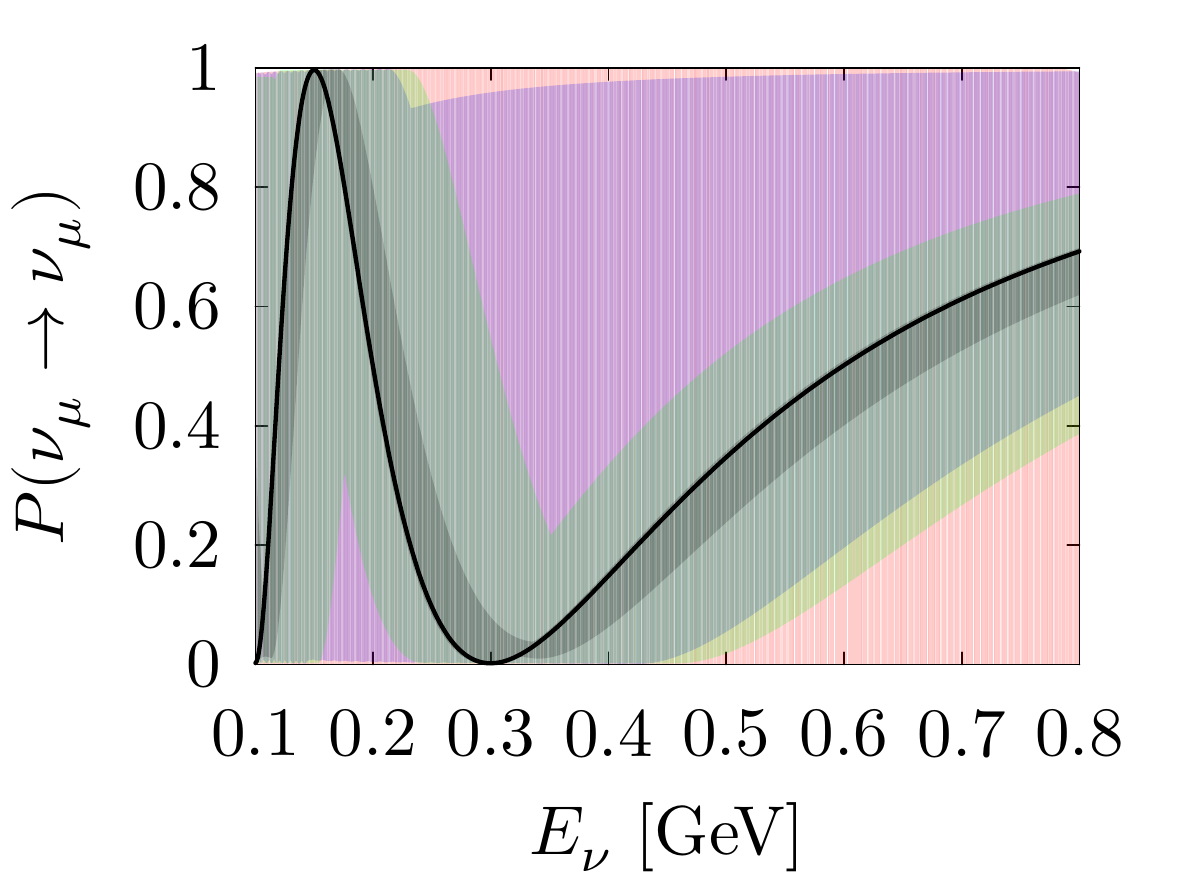}
 \includegraphics[width=0.45\textwidth]{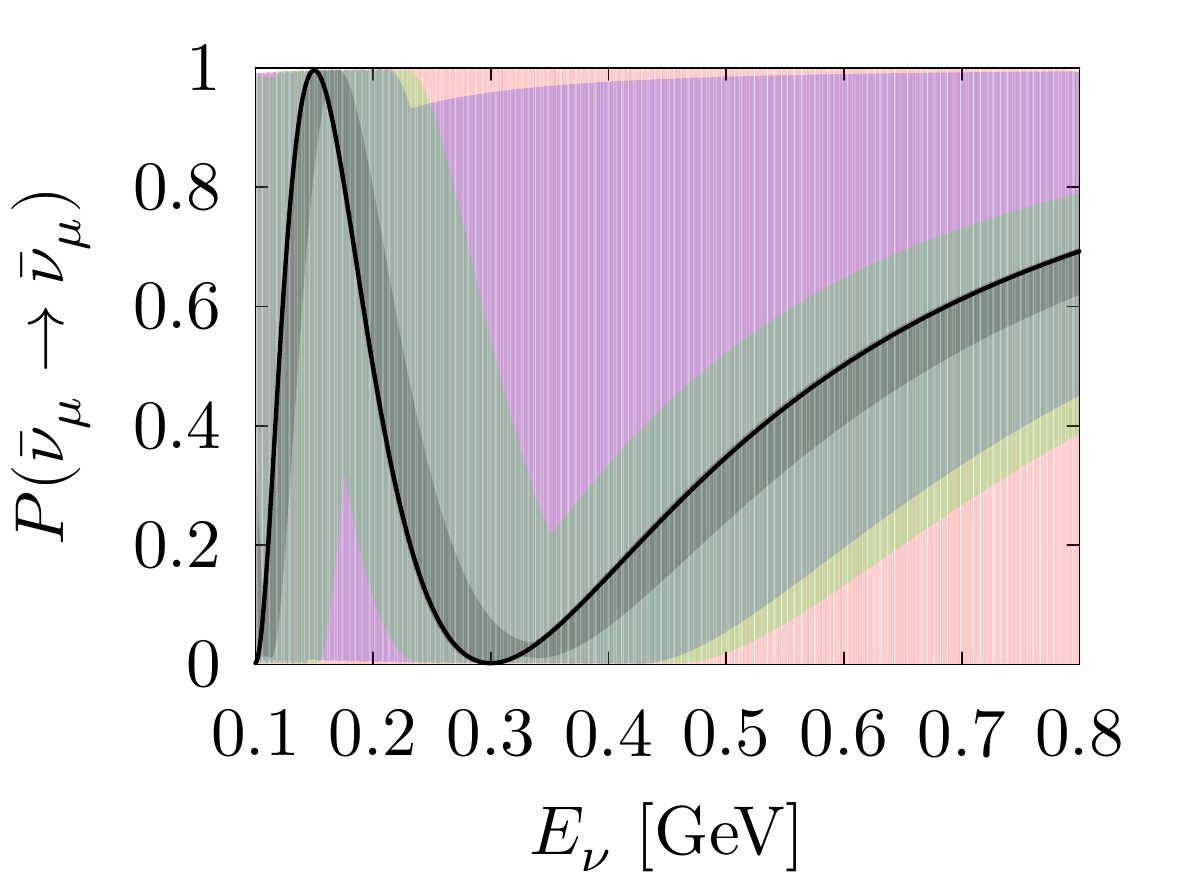}
 \caption{\label{fig:prob_dis}.The impact of the probability for varying each model parameters within $3\sigma$ uncertainty predicted with NuFit4.0 result Eq.~(\ref{eq:3sigma_nufit4}):  $-5.475<x<-3.37$ (red band),~$0.455<\eta/\pi<1.545$ (dark grey band), ~$0.204<r<0.606$ (blue band),~$3.343<m_a/\text{meV}<4.597$ (green band). We also show the probability for the best fit (B.F.) Table~\ref{tab:model_fit} in the black curve:$(x,~\eta,~r,~M_a)=(-3.65,~1.13\pi,~0.511,~3.71~\text{meV})$. The upper left (right) panel is for $P(\nu_e\rightarrow\nu_e)$ ($P(\bar{\nu}_e\rightarrow\bar{\nu}_e)$), while the lower left (right) panel is for $P(\nu_\mu\rightarrow\nu_\mu)$ ($P(\bar{\nu}_\mu\rightarrow\bar{\nu}_\mu)$).}
\end{figure}

\begin{figure}[!h]
 \centering
\includegraphics[width=0.45\textwidth]{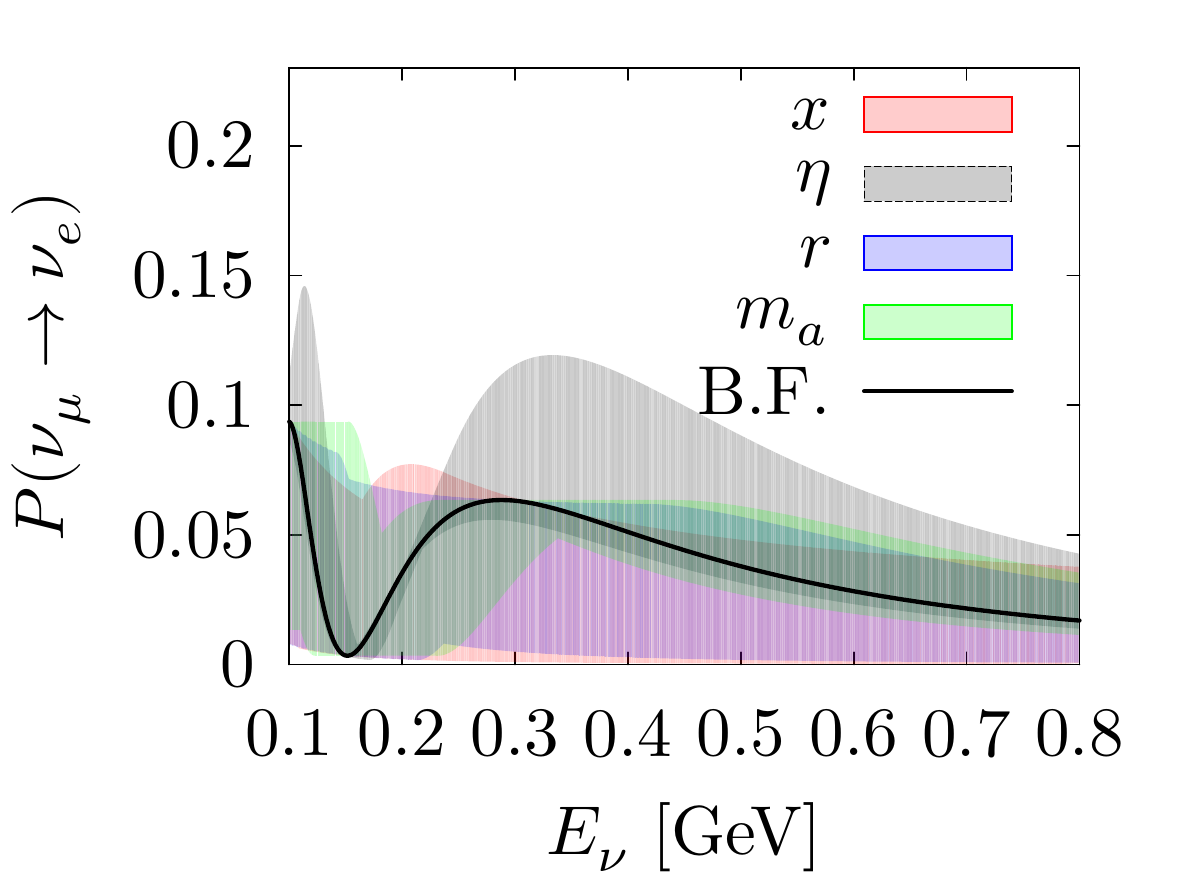}
 \includegraphics[width=0.45\textwidth]{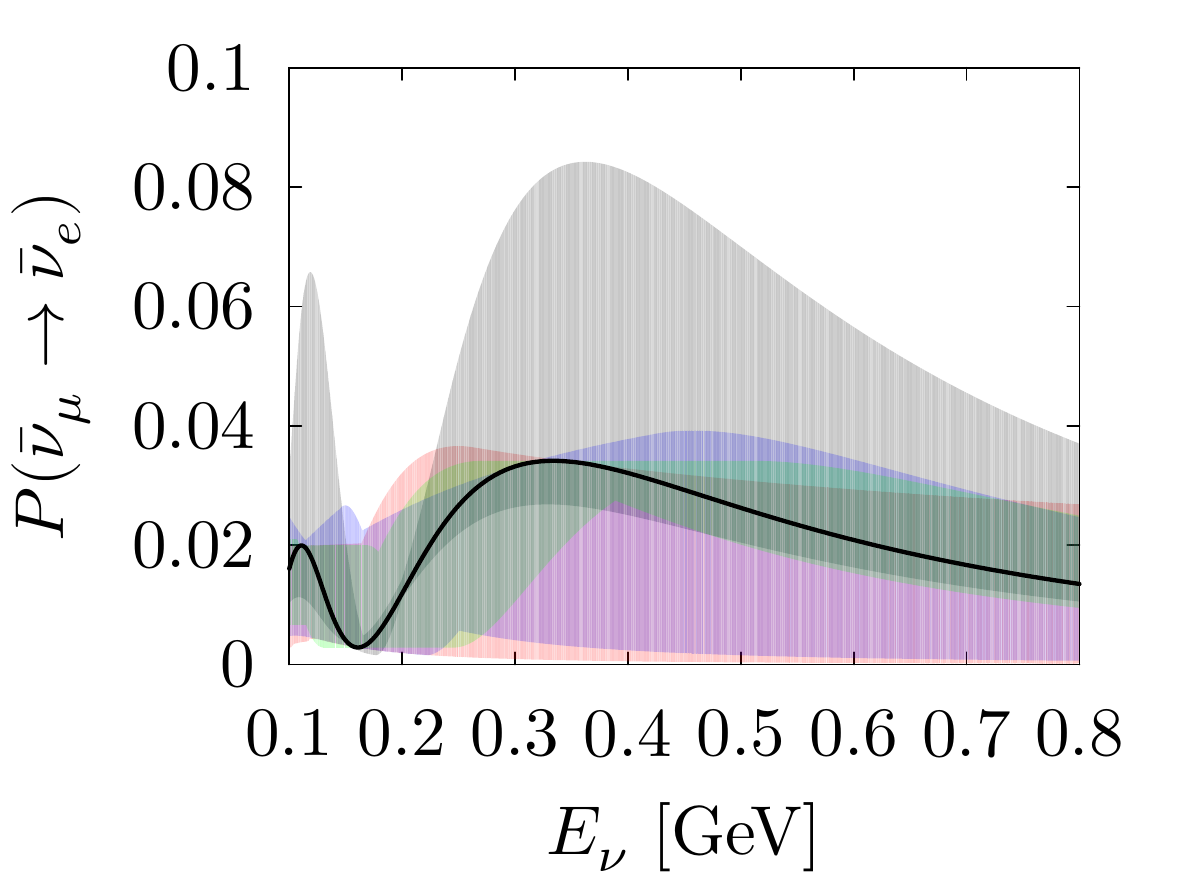}$~~~~~~$\\
 \includegraphics[width=0.45\textwidth]{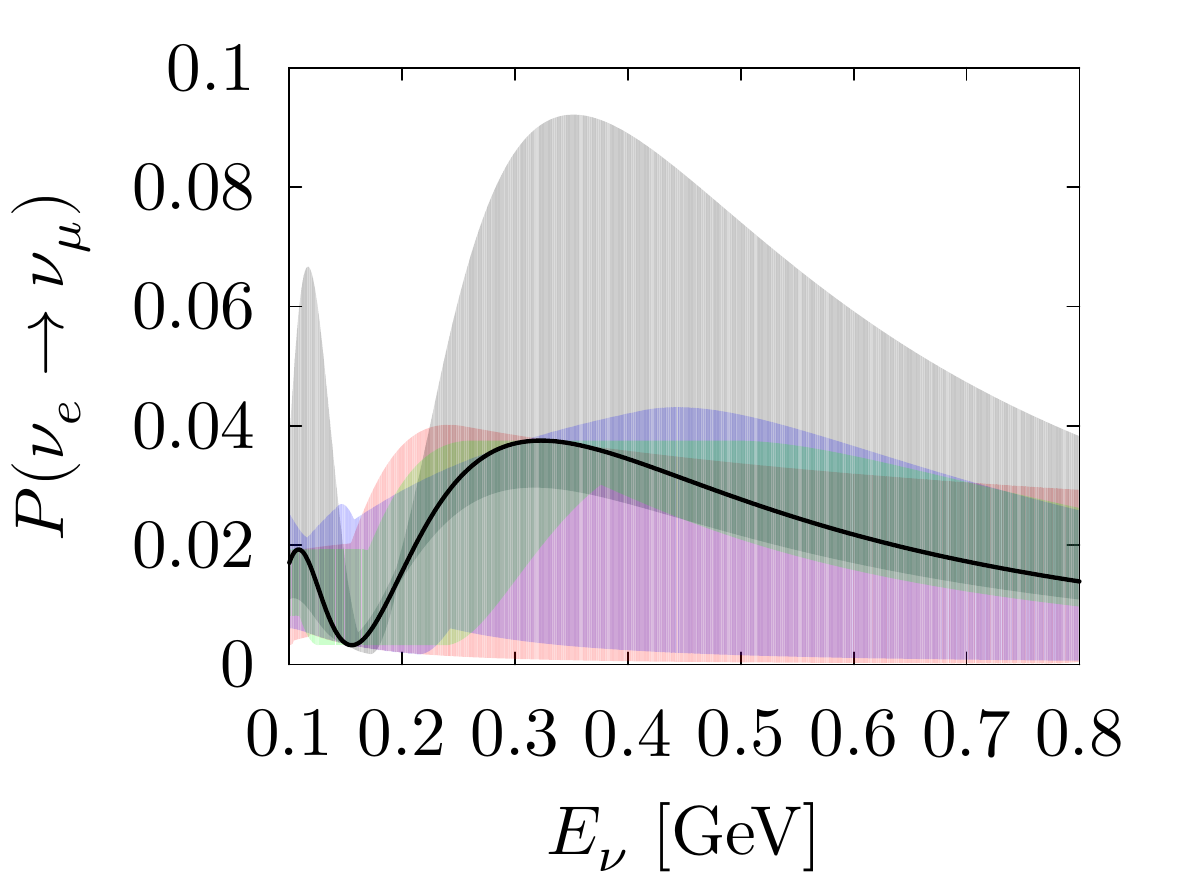}
 \includegraphics[width=0.45\textwidth]{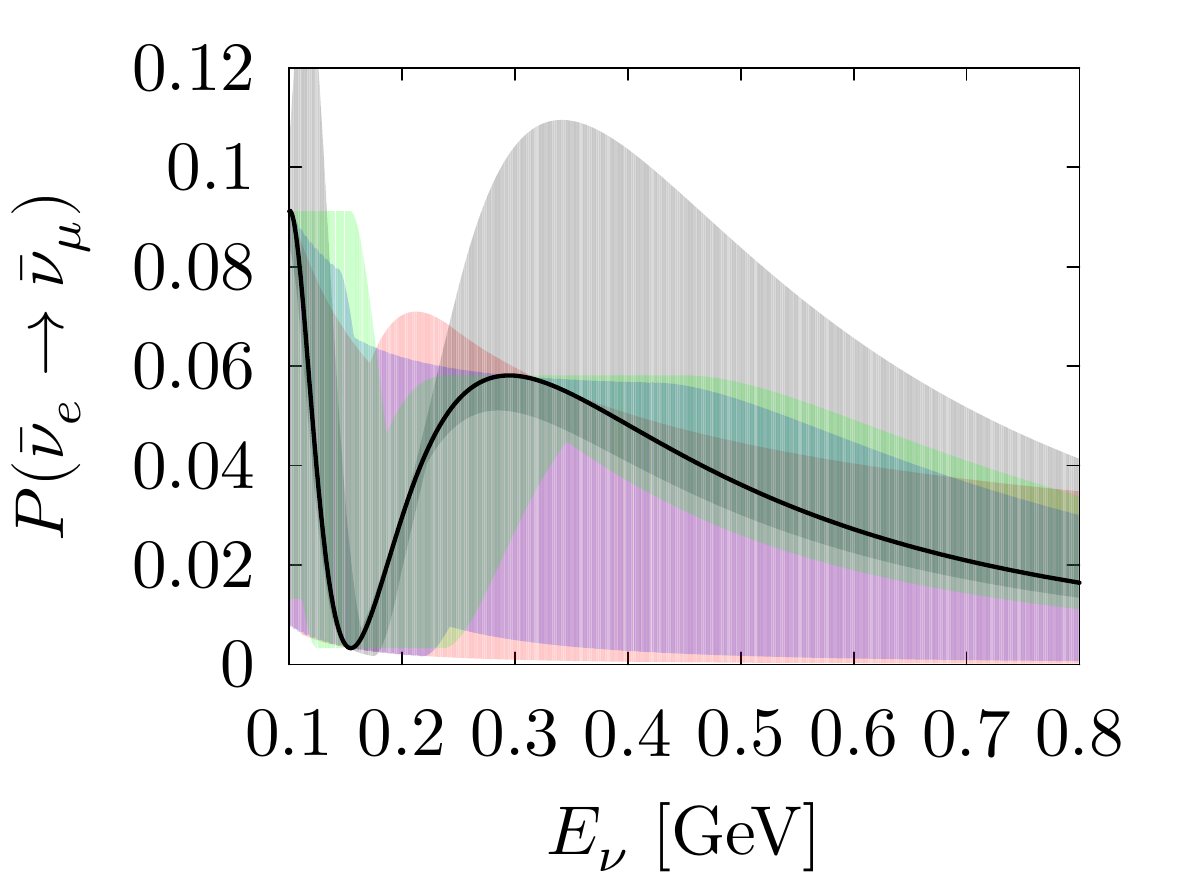}
 \caption{\label{fig:prob_app}.The impact of the probability for varying each model parameters within $3\sigma$ uncertainty predicted with NuFit4.0 result Eq.~(\ref{eq:3sigma_nufit4}):  $-5.475<x<-3.37$ (red band),~$0.455<\eta/\pi<1.545$ (dark grey band), ~$0.204<r<0.606$ (blue band),~$3.343<m_a/\text{meV}<4.597$ (green band). We also show the probability for the best fit (B.F.) Table~\ref{tab:model_fit} in the black curve: $(x,~\eta,~r,~m_a)=(-3.65,~1.13\pi,~0.511,~3.71~\text{meV})$. The upper left (right) panel is for $P(\nu_\mu\rightarrow\nu_e)$ ($P(\bar{\nu}_\mu\rightarrow\bar{\nu}_e)$), while the lower left (right) panel is for $P(\nu_e\rightarrow\nu_\mu)$ ($P(\bar{\nu}_e\rightarrow\bar{\nu}_\mu)$).}
\end{figure}

In Figs.~\ref{fig:prob_dis} and \ref{fig:prob_app}, we present the variation of probabilities for MOMENT with the $3\sigma$ uncertainty for model parameters in terms of NuFit4.0 results given in Eq.~(\ref{eq:3sigma_nufit4}). We also show the probability with the best fit values as the input Table~\ref{tab:model_fit}.
 In Fig.~\ref{fig:prob_dis}, we see the variation of $\nu_\mu$ and $\bar{\nu}_\mu$ disappearance channels is much larger than those in the electron neutrino disappearance channels. As a result, $\nu_\mu$ and $\bar{\nu}_\mu$ disappearance channels are two most dominating channels for the TDLS model. In the lower two panels, we see the variation of $x$ in the model has the largest impact, covering the range from $0$ to $1$ for the probability within $0.1~\text{GeV}\le E_\nu\le 0.8~\text{GeV}$. The second largest effect comes from the model parameter $r$. It also ranges from $0$ to $1$, yet the trend is different. For the higher energy ($E_\nu>0.45$ GeV), the lower bound of the probability is getting larger, and it is $\sim 0.45$ at $E_\nu=0.8$ GeV for both channels. For the model parameter $m_a$, the probability is changing with $\Delta P\sim 0.2$ along with the probability for the best fit value in Table~\ref{tab:model_fit}. The similar feature is seen for the parameter $\eta$; yet the variation of probability is smaller $\Delta P\sim 0.05$. It seems that $\eta$ is the distinctive parameter not to be measured by $\nu_\mu$ and $\bar{\nu}_\mu$ disappearance channels as easily as the other three model parameters. Eventually, we find that $\nu_e$ and $\bar{\nu}_e$ disappearance channels are more sensitive to the variation of $\eta$ than the other parameters, where $\Delta P$ can approach $\sim 0.1$ around the first minimum $E_\nu\sim0.3$ GeV.

In Fig.~\ref{fig:prob_app}, we show variations of $P(\nu_\mu\rightarrow\nu_e)$, $P(\bar{\nu}_\mu\rightarrow\bar{\nu}_e)$, $P(\nu_e\rightarrow\nu_\mu)$, and $P(\bar{\nu}_e\rightarrow\bar{\nu}_\mu)$. The behaviours in four panels are almost the same. The largest variation is given by the impact of $\eta$: $\Delta P\sim0.06$ around the first maximum $E_\nu\sim0.3$ GeV for all panels. The impact of model parameters $x$ and $r$ can reduce the lower bound significantly in the probability plane. From the first minimum to $8$ GeV, the lower bound of probability can even reach $0$. 
For both parameters, the variation of probability is around $\Delta P\sim0.03$. The variation for $m_a$ is the smallest around $0.01$.

\begin{figure}[!h]
 \centering
\includegraphics[width=0.45\textwidth]{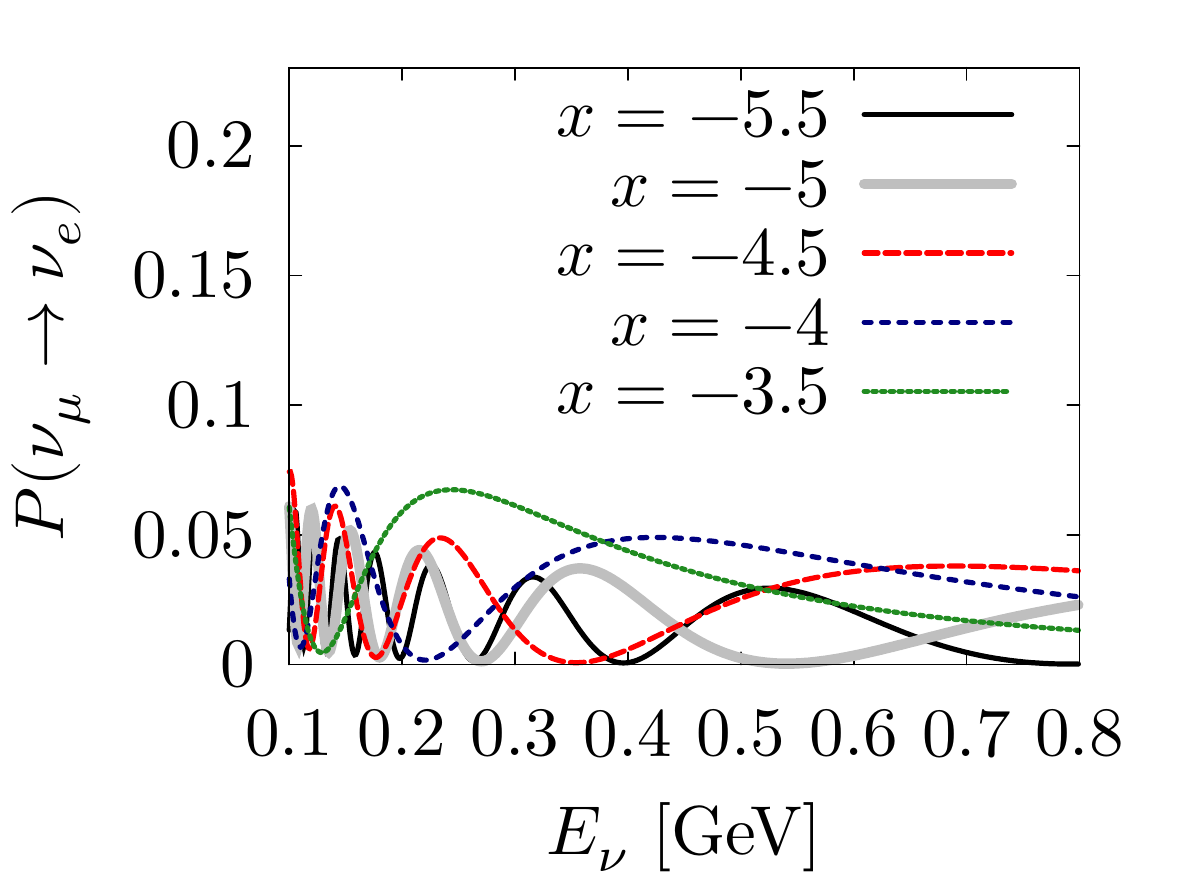}
 \includegraphics[width=0.45\textwidth]{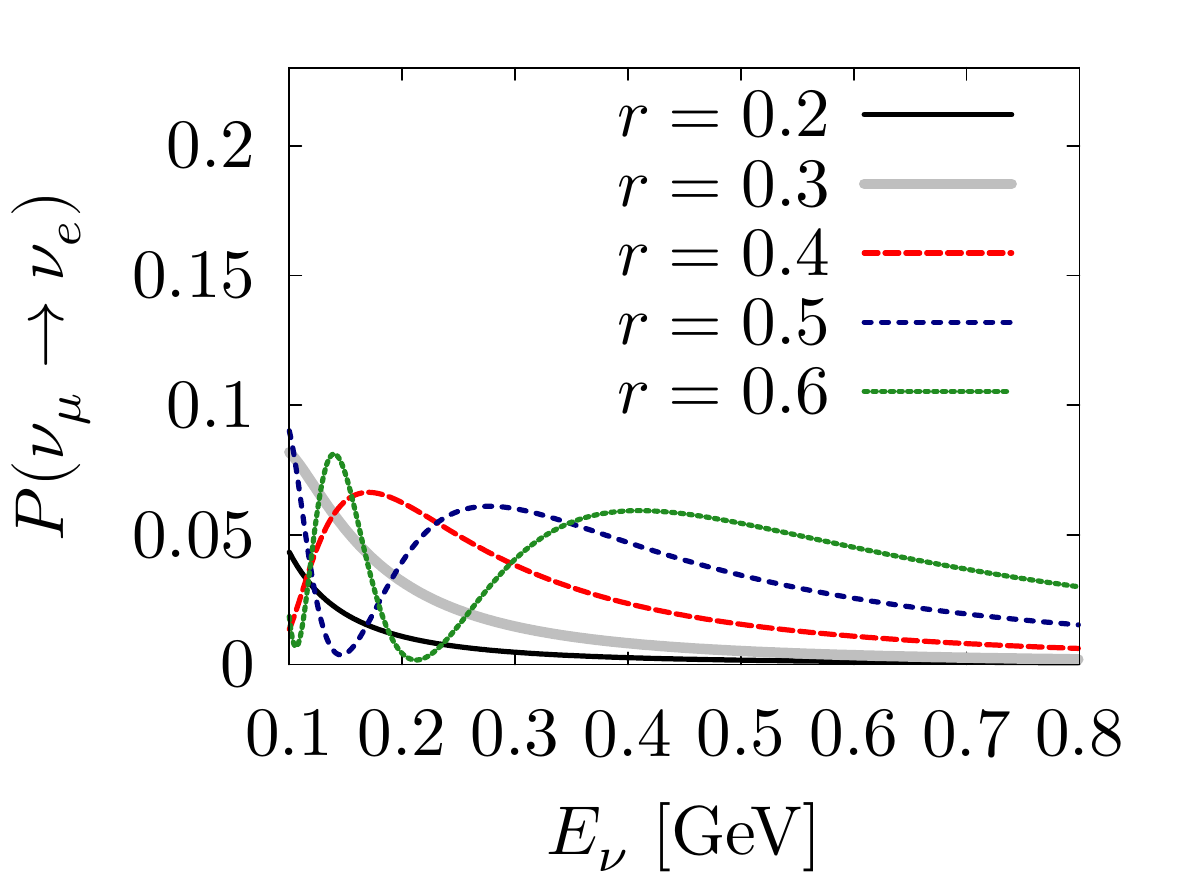}
 \caption{\label{fig:prob_app_xr}.The impact of the probability $P(\nu_\mu\rightarrow\nu_e)$ for varying value for $x$ (left) and $r$ (right). Except for the varied one, the other parameters are used according to Table~\ref{tab:model_fit} in the black curve: $(x,~\eta,~r,~m_a)=(-3.65,~1.13\pi,~0.511,~3.71~\text{meV})$. In the left (right) panel, the black, thick grey, red-dashed, blue-short-dashed, and green-dotted curves are $x=-5.5,~-5,~-4.5,~-4,~-3.5$ ($r=0.2,~0.3,~0.4,~0.5,~0.6$), respectively.}
\end{figure}

We observed that the lower limits reach $0$ in a wide range of $E_\nu$ for most of channels, except $\nu_e$ and $\bar{\nu}_e$ disappearance ones. This happens when we varying the values of $x$ and $r$. The reason for this feature is that the oscillation minimum moves in wide range of $E_\nu$ with $x$ or $r$, as we see in Fig.~\ref{fig:prob_app_xr}, in which we use $P(\nu_\mu\rightarrow\nu_e)$ as an example. We vary $x$ from $-5.5$ to $-3.5$ (left panel), and vary $r$ from $0.2$ to $0.6$ (right panel). The result demonstrates that the horizontal shift of the minimum makes the lower limit of the band to be $0$ in a wide $E_\nu$ region.

To sum up, we see that $\nu_\mu$ and $\bar{\nu}_\mu$ disappearance channels are the most important channels to constrain TDLS models, especially for $x$, $r$ and $m_a$. However, the other six channels can provide information for $\eta$. Thanks to the multiple channel features, MOMENT can be used to study TDLS models and can even measure model parameters precisely.

%%%%%%%%%%%%%%%%%%%%%%%%%%%%%%%%%%%%%%%%
%%%%%%%%%%%%%%%%%%%%%%%%%%%%%%%%%%%%%%%%
%%%%%%%%%%%%%%%%%RESULTS%%%%%%%%%%%%%%%%%%
%%%%%%%%%%%%%%%%%%%%%%%%%%%%%%%%%%%%%%%%
%%%%%%%%%%%%%%%%%%%%%%%%%%%%%%%%%%%%%%%%

\section{Results}\label{sec:results}
In this section, we present physics potentials of MOMENT on the TDLS model. We firstly predict the exclusion limit for this model in different scenarios. We will see that $\theta_{23}$ and $\delta$ are key parameters to exclude TDLS models. Then, we study how MOMENT data can be used to constrain model parameters. We will see model-parameter degeneracies due to the poor measurement of $\theta_{12}$. We also project the $\Delta\chi^2$ to the standard neutrino mixing parameter space from the model parameter space. This shows an interesting correlation and demonstrate the goodness of fit in the analysis of simulated data.

%%%%%%%%%%%%%%%%%%%%%%%%%%%%%%%%%%%%%%%%%%%%%%%%%%%%%%%
\subsection{Model Exclusion}\label{sec:exclusion}

To give the model exclusion curves, we study the minimum of $\chi^2$ value for the TDLS with a given set of true values for the standard oscillation parameters (three mixing angles, two mass-square splittings, and a Dirac CP phase), and define the statistical quantity $\chi^2_{ex.}$ as follows:
\begin{equation}\label{eq:sum_rule}
\chi^2_{ex.}=\sum_i\min_{\overrightarrow{\mathcal{M}}}\chi^2(\mu_i(\overrightarrow{\mathcal{M}}),n_i(\overrightarrow{\mathcal{O}}_{true})).
\end{equation}
We adopt Wilk's theorem \cite{Wilks:1938dza}. When comparing nested models, the $\Delta \chi^2$ test statistics is a random variable asymptotically distributed according to the $\chi^2$-distribution with the number of degrees of freedom, which is equal to the difference in the number of free model parameters.

\begin{figure}[!h]%
\centering
\includegraphics[width=2.7in]{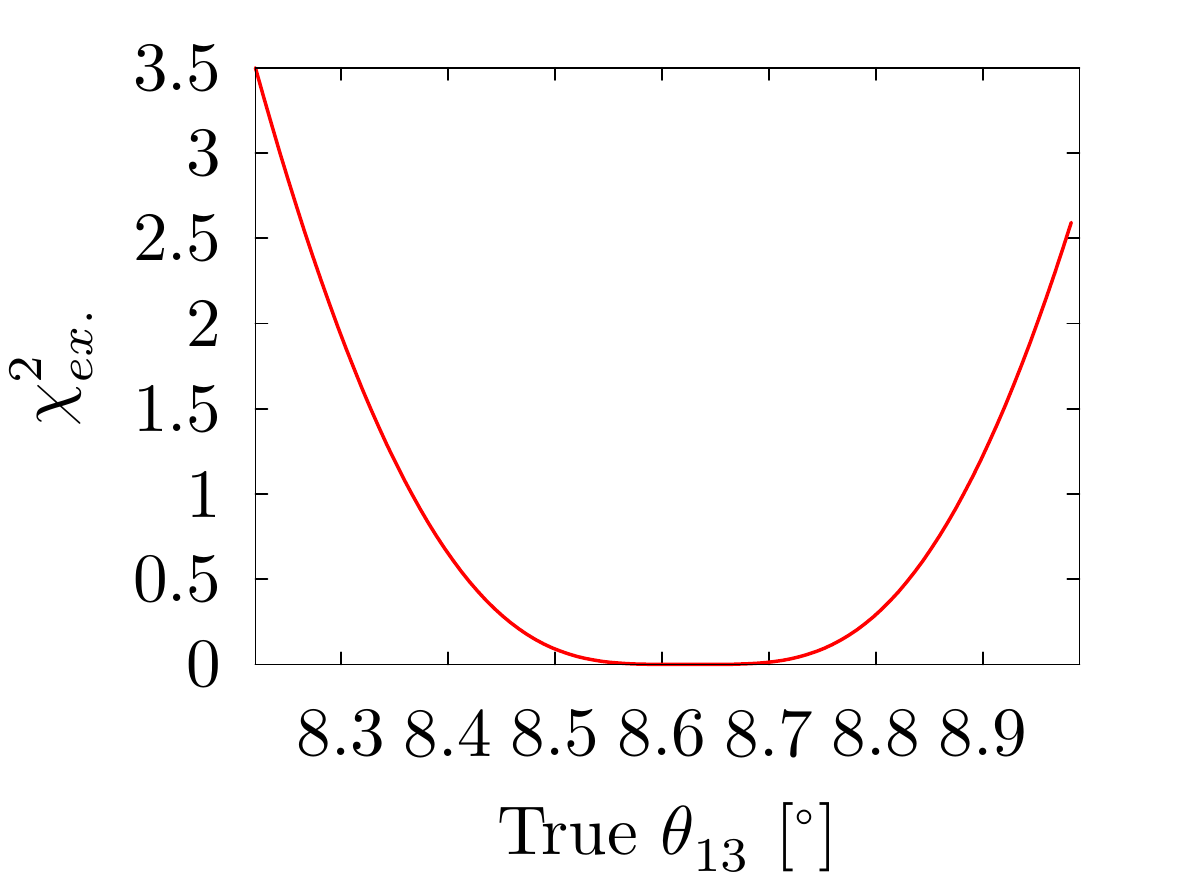}
\includegraphics[width=2.7in]{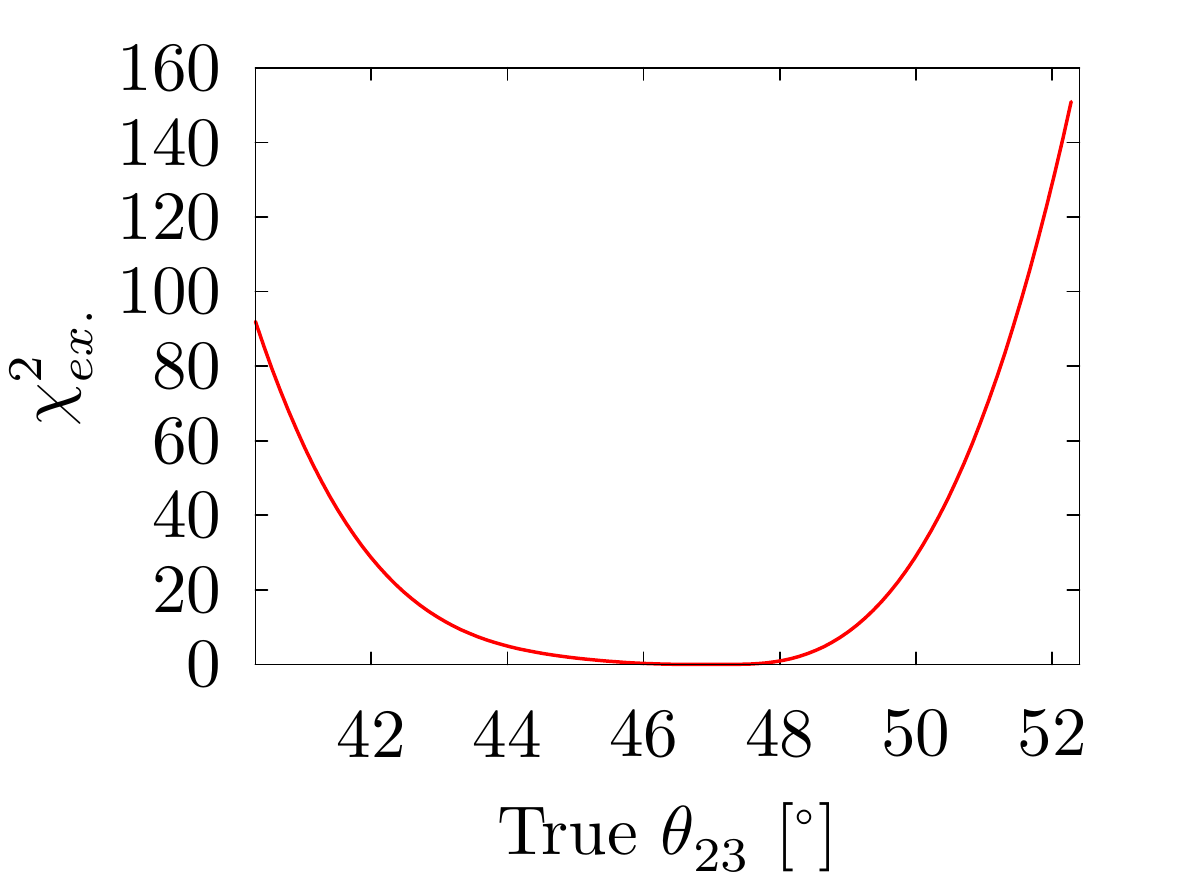}\\
\includegraphics[width=2.7in]{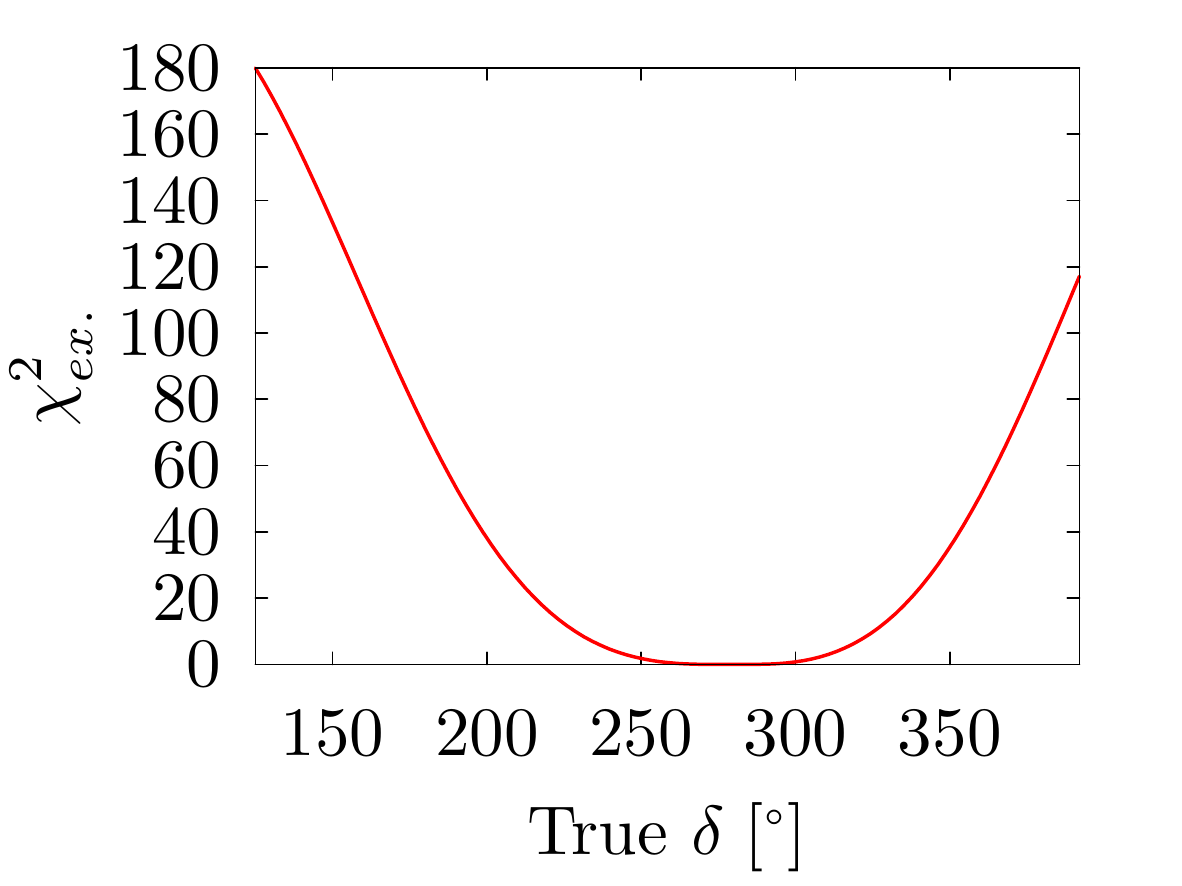}
\includegraphics[width=2.7in]{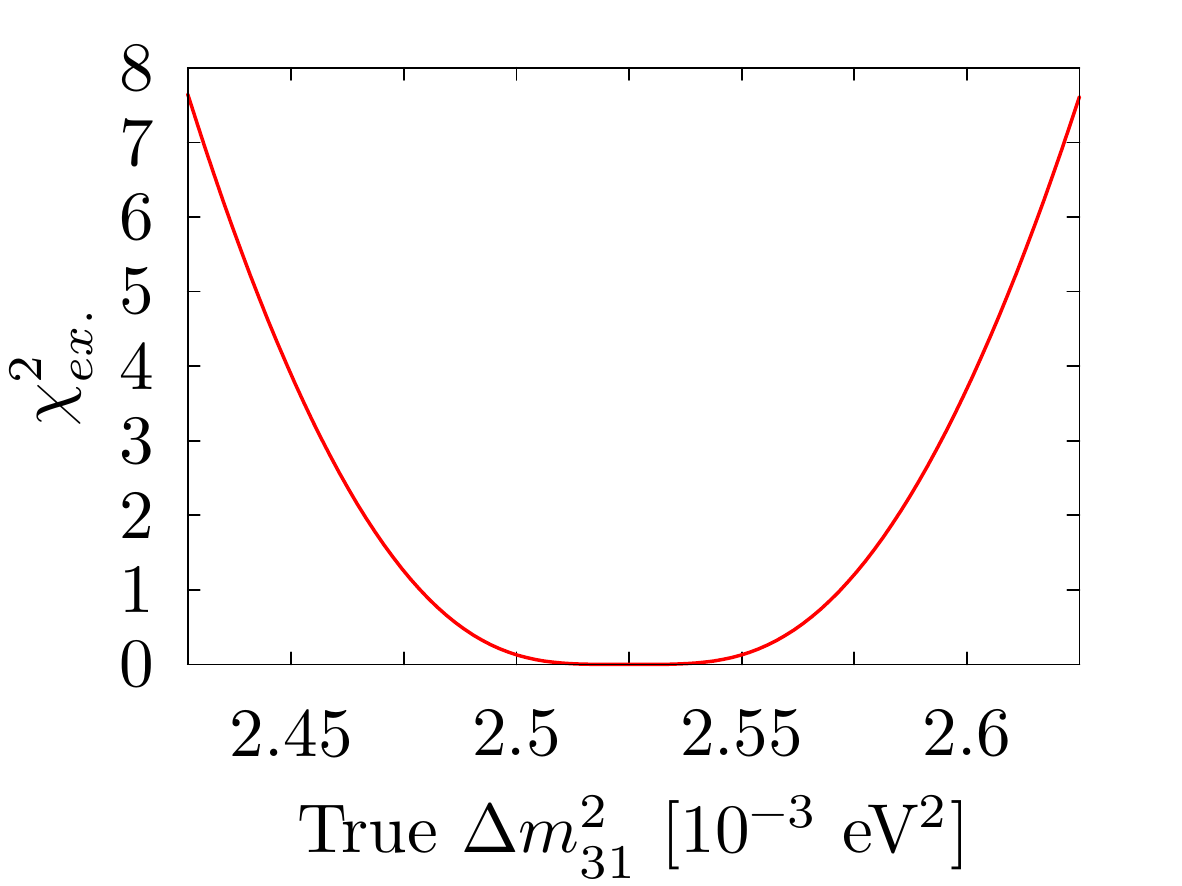}
\caption{The $\chi^2_{ex.}$ value for tri-direct littlest seesaw model for $\theta_{13}$, $\theta_{23}$, $\delta$ and $\Delta m_{31}^2$. The range for each parameter is taken according to the $3\sigma$ uncertainty in NuFit4.0 results.}%
\label{fig:SR_1D}
\end{figure}

\begin{figure}[!h]
 \flushleft
\includegraphics[width=0.32\textwidth]{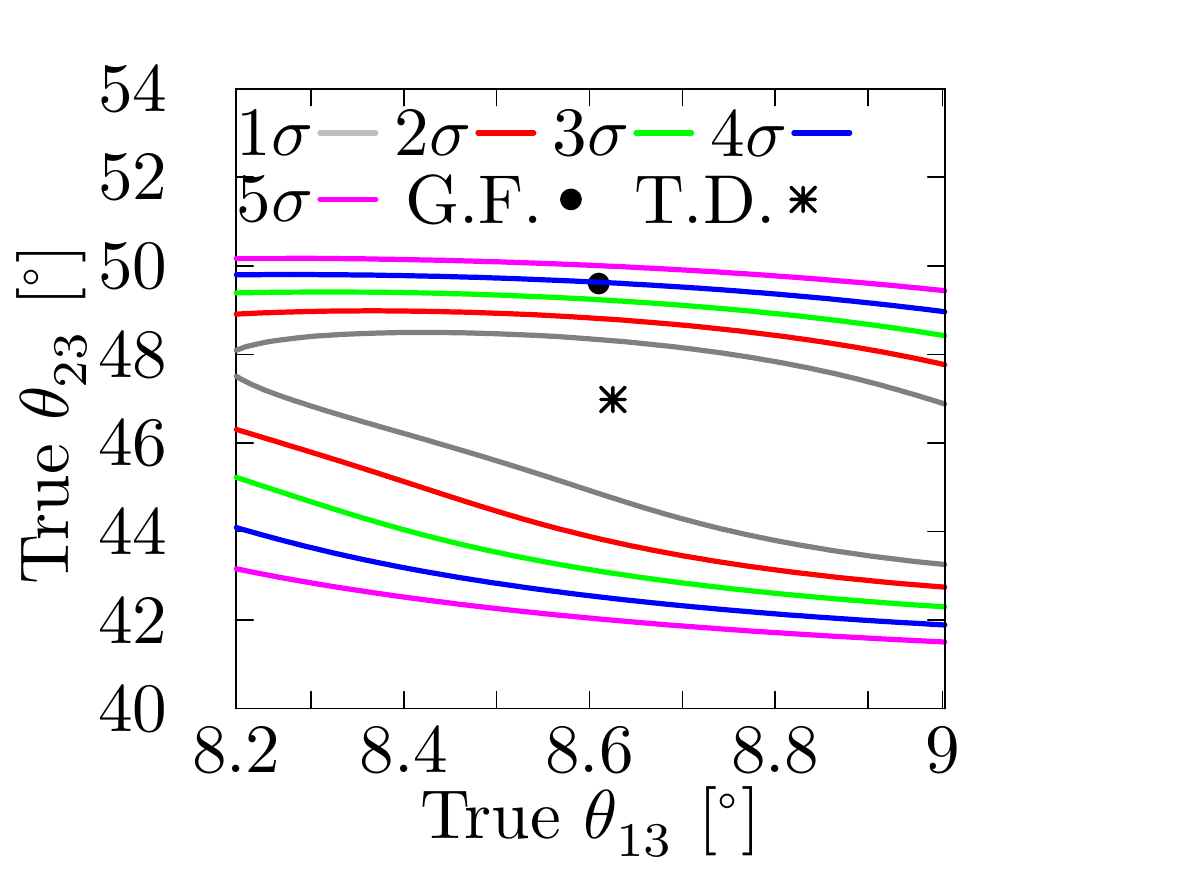}$~~~~~~$\\
\includegraphics[width=0.32\textwidth]{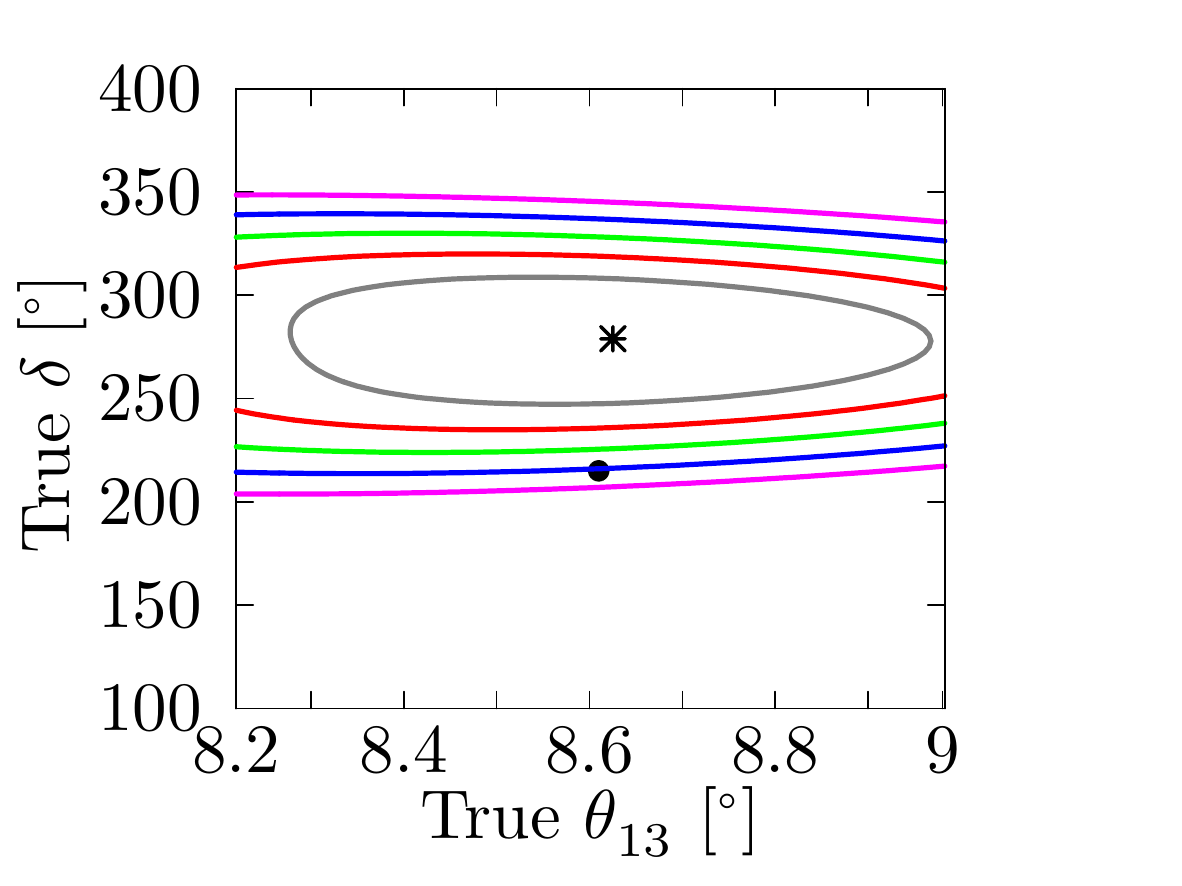}
\includegraphics[width=0.32\textwidth]{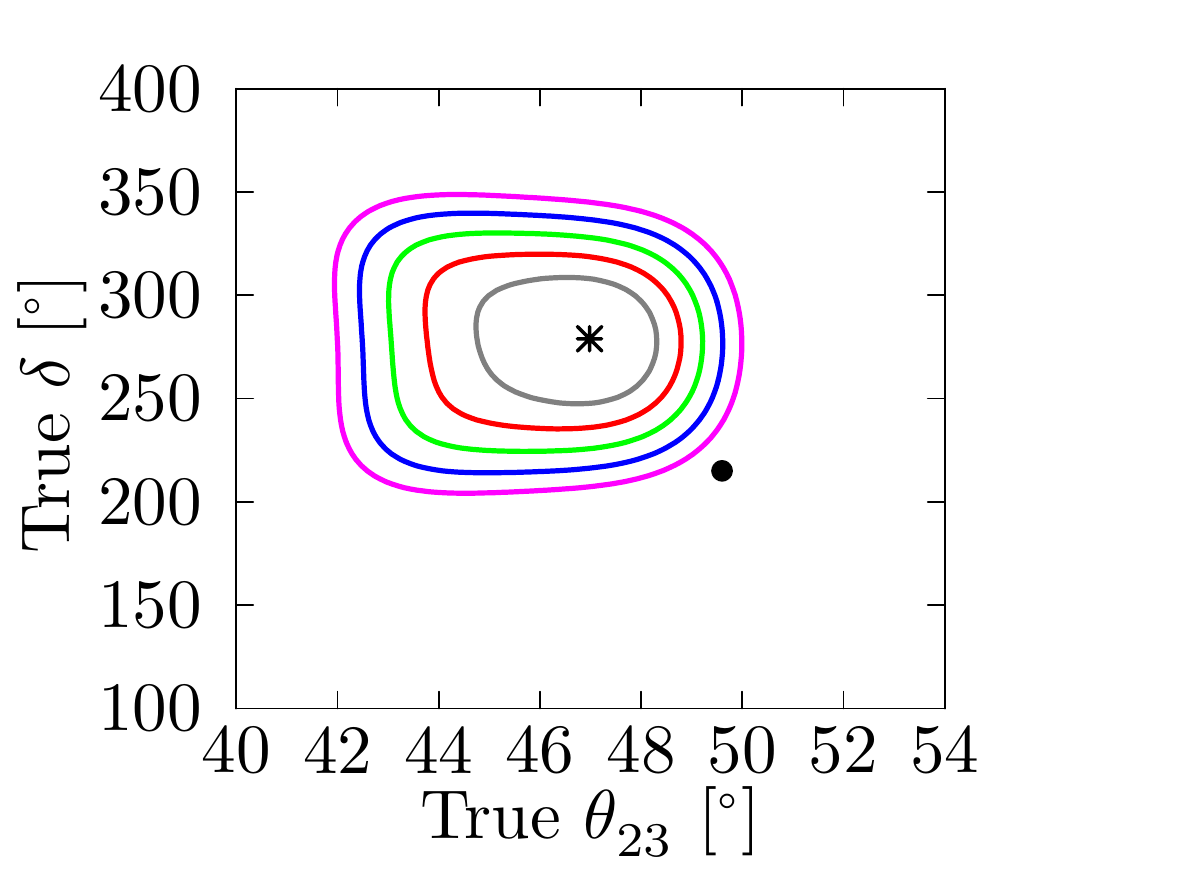}$~~~~~~$\\
\includegraphics[width=0.32\textwidth]{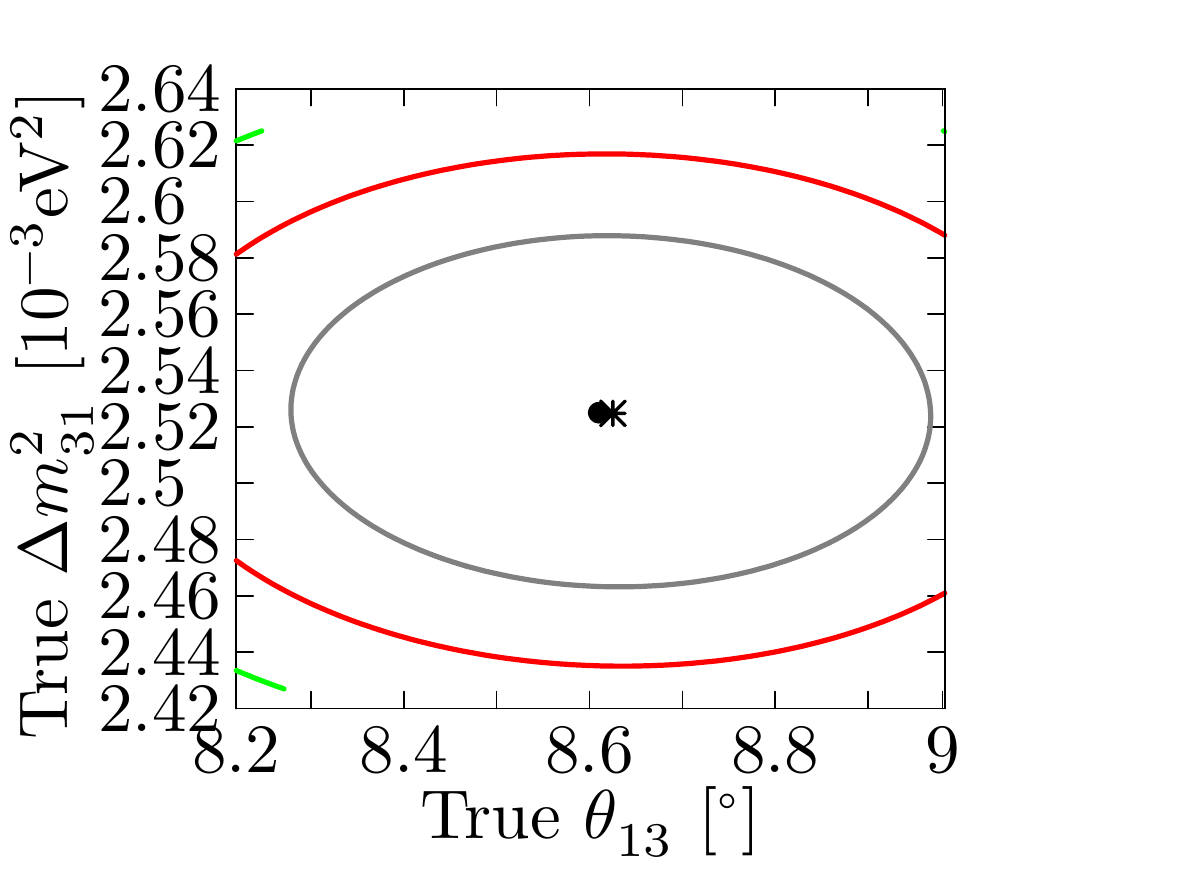}
\includegraphics[width=0.32\textwidth]{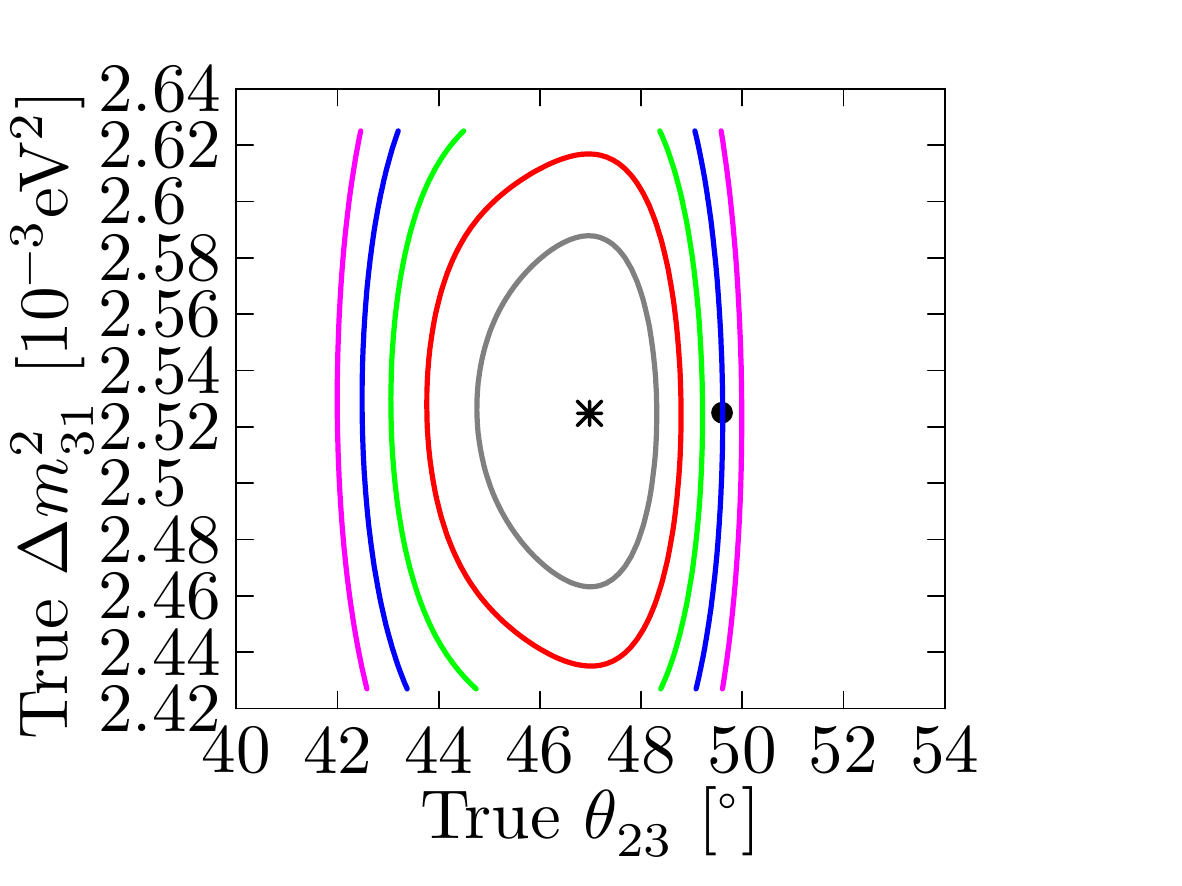}
\includegraphics[width=0.32\textwidth]{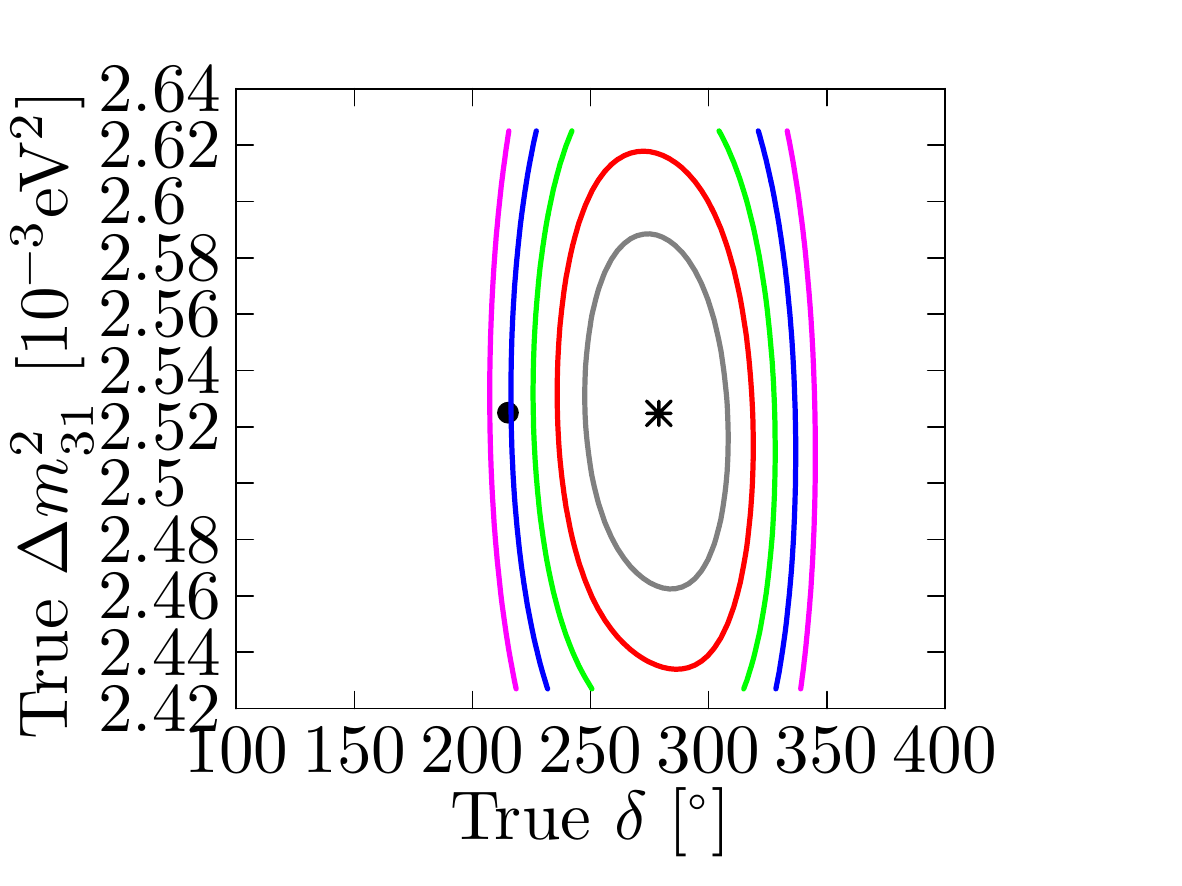}
 \caption{\label{fig:SR_2D}The 2-D exclusion contour for tri-direct littlest seesaw model on the plane of any two true standard parameters, from $1\sigma$ to $5\sigma$. The range for each parameter is taken according to the $3\sigma$ uncertainty in NuFit4.0 results. The black dot denotes the best fit of NuFit4.0 results ($(\theta_{12},~\theta_{13},~\theta_{23},~\delta,~\Delta m_{21}^2,~\Delta m_{31}^2)=(33.82^\circ,~8.61^\circ,~49.6^\circ,~215^\circ,~7.39\times10^{-5}~\text{eV}^2,~2.525\times10^{-3}~\text{eV}^2)$), while the star is the prediction by the tri-direct littlest seesaw model with NuFit4.0 results ($(\theta_{12},~\theta_{13},~\theta_{23},~\delta,~\Delta m_{21}^2,~\Delta m_{31}^2)\sim(36.25^\circ,~8.63^\circ,~47^\circ,~279^\circ,~7.39\times10^{-5}~\text{eV}^2,~2.525\times10^{-3}~\text{eV}^2)$).}
\end{figure}

We present our result in Figs.~\ref{fig:SR_1D} and \ref{fig:SR_2D}. In these figures, we vary true values for each one or two of standard oscillation parameters, while the other standard oscillation parameters are fixed at the TDLS predictions $(\theta_{12},~\theta_{13},~\theta_{23},~\delta,~\Delta m_{21}^2,~\Delta m_{31}^2)\sim(36.25^\circ,~8.63^\circ,~47^\circ,~279^\circ,~7.39\times10^{-5}~\text{eV}^2,~2.525\times10^{-3}~\text{eV}^2)$. As we do not see any impact on $\theta_{12}$ and $\Delta m_{21}^2$, we will simply ignore them in our discussion from now on.

In Fig.~\ref{fig:SR_1D}, we show the $\chi^2_{ex.}$ values against various true values for $\theta_{13}$ (upper-left), $\theta_{23}$ (upper-right), $\delta$ (lower-left), $\Delta m_{31}^2$ (lower-right). The range we show is given by the $3\sigma$ uncertainty in the NuFit4.0. Strikingly, we see very high exclusion levels for $\theta_{23}$ and $\delta$; for $\theta_{23}$ ($\delta$), $\chi^2_{ex.}$ can climb up to $\sim160$ ($\sim120$) at the upper bound, and reach $\sim90$ ($\sim180$) at the lower bound. For $\Delta m_{31}^2$, the exclusion level $\chi^2_{ex.}$ at both bounds is close to $8$. The worst one among these four parameters is $\theta_{13}$, and it cannot even reach $2\sigma$ exclusion level at the $3\sigma$ uncertainty of NuFit4.0.

In Fig.~\ref{fig:SR_2D}, we show 2-dimension contours at $1\sigma$ (gray), $2\sigma$ (red), $3\sigma$ (green), $4\sigma$ (blue), and $5\sigma$ (magenta) on a combination of two parameters from $\theta_{13}$, $\theta_{23}$, $\delta$, and $\Delta m_{31}^2$. The range we show is the $3\sigma$ uncertainty in NuFit4.0. In all panels, the black dot denotes the best fit of NuFit4.0 results ($(\theta_{12},~\theta_{13},~\theta_{23},~\delta,~\Delta m_{21}^2,~\Delta m_{31}^2)=(33.82^\circ,~8.61^\circ,~49.6^\circ,~215^\circ,~7.39\times10^{-5}~\text{eV}^2,~2.525\times10^{-3}~\text{eV}^2)$), while the star is the prediction by the tri-direct littlest seesaw model with NuFit4.0 results ($(\theta_{12},~\theta_{13},~\theta_{23},~\delta,~\Delta m_{21}^2,~\Delta m_{31}^2)\sim(36.25^\circ,~8.63^\circ,~47.^\circ,~279^\circ,~7.39\times10^{-5}~\text{eV}^2,~2.525\times10^{-3}~\text{eV}^2)$). Though we do not see any correlations, we find that the black dot is outside of $5\sigma$ contour on the $\theta_{23}$-$\delta$ plane. This tells us that the measurement of $\theta_{23}$ and $\delta$ for MOMENT can exclude  the TDLS over $5\sigma$ if NuFit4.0 results are confirmed.
%%%%%%%%%%%%%%%%%%%%%%%%%%%%%%%%%%%%%%%%%%%%%%%%%%%%%%%
\subsection{Model parameter constraint}\label{sec:Model_Cons}

\begin{figure}[!h]%
\centering
\includegraphics[width=2.7in]{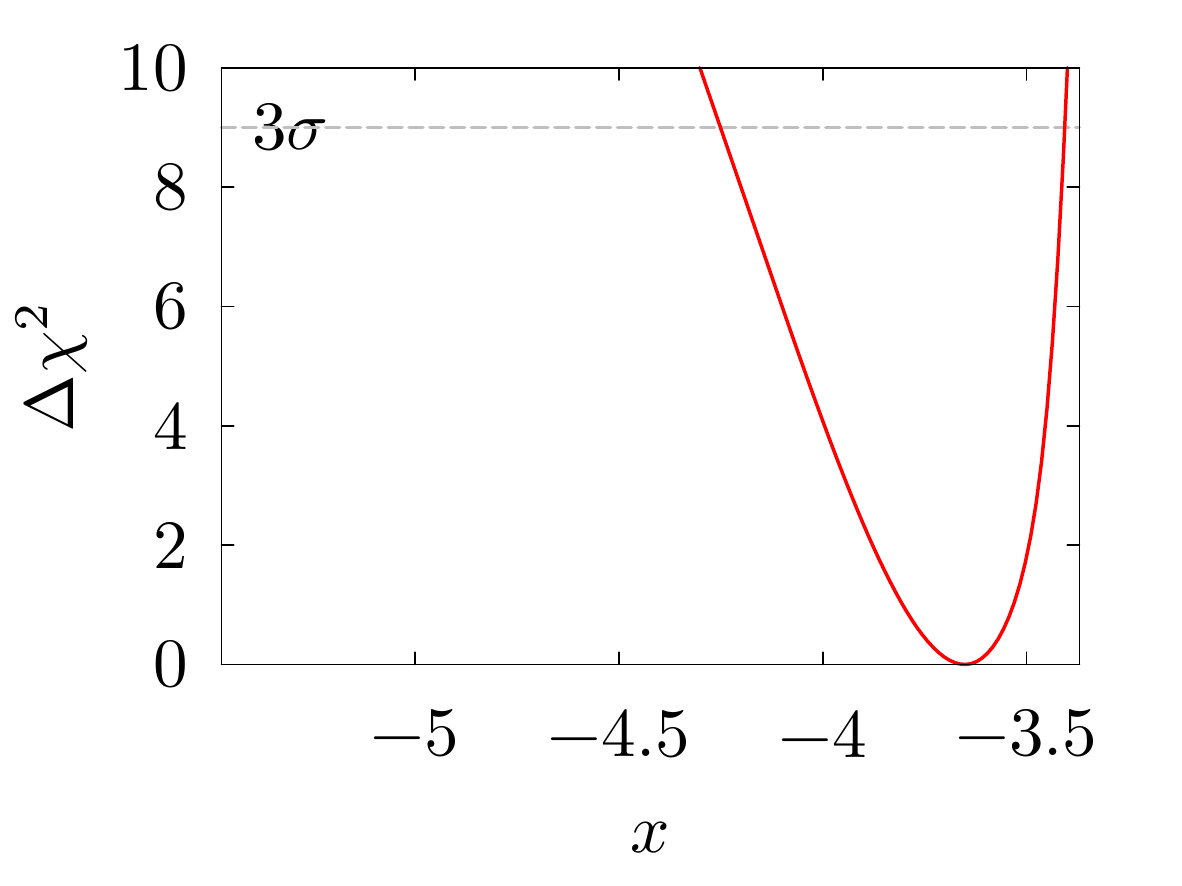}
\includegraphics[width=2.7in]{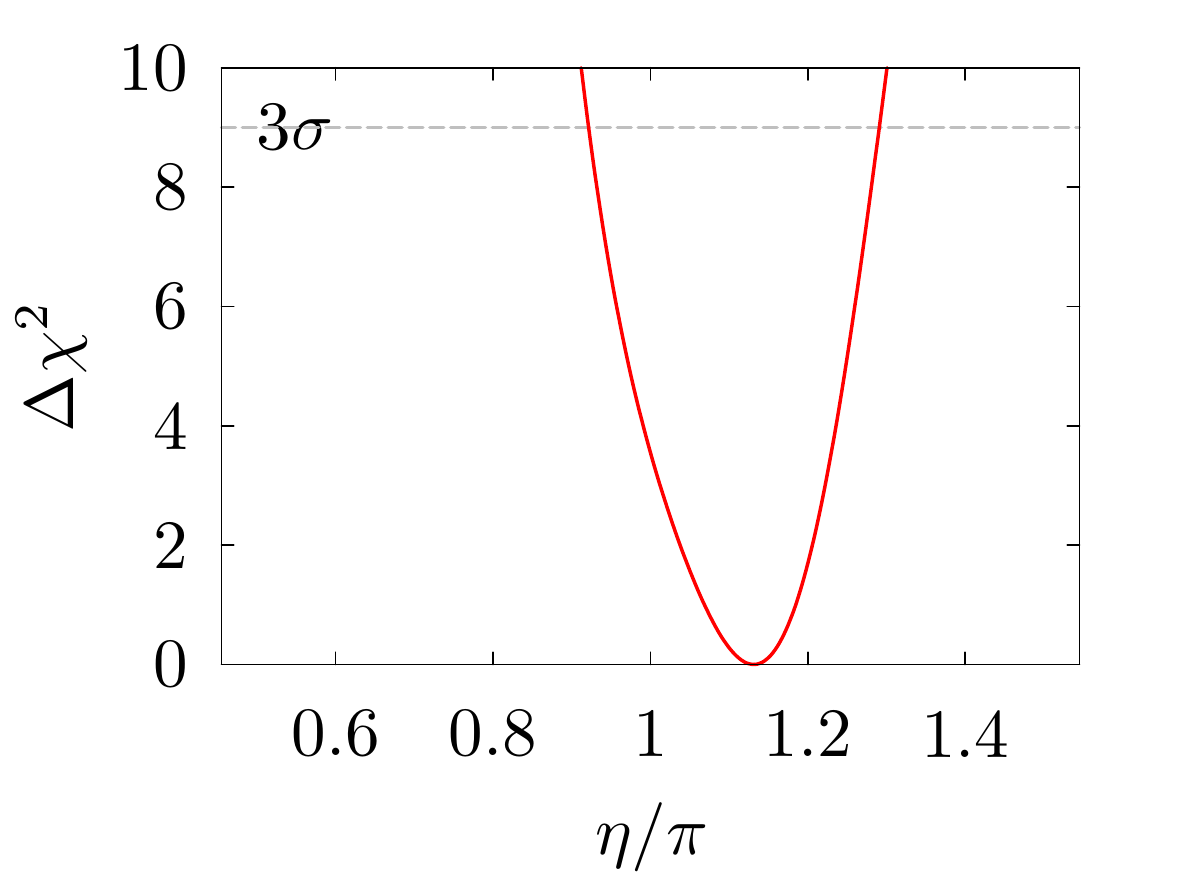}\\
\includegraphics[width=2.7in]{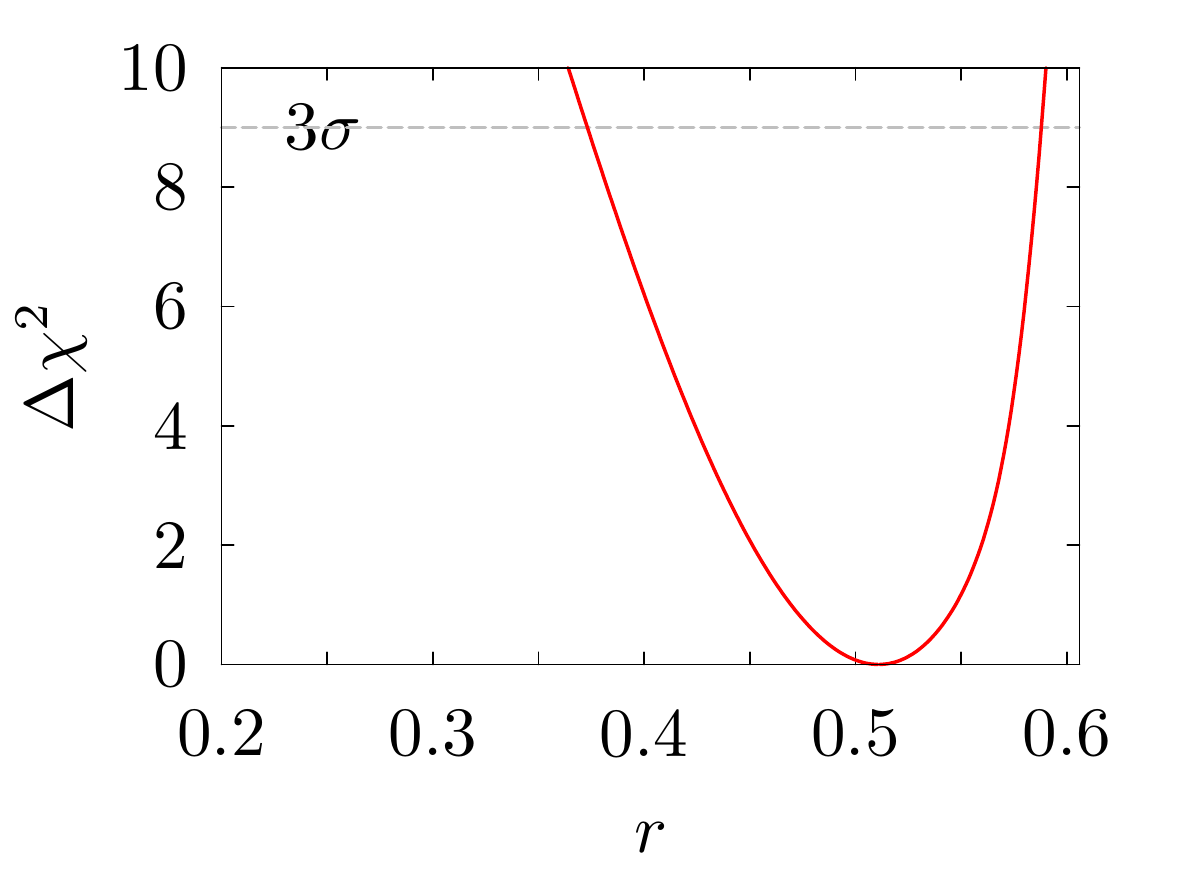}
\includegraphics[width=2.7in]{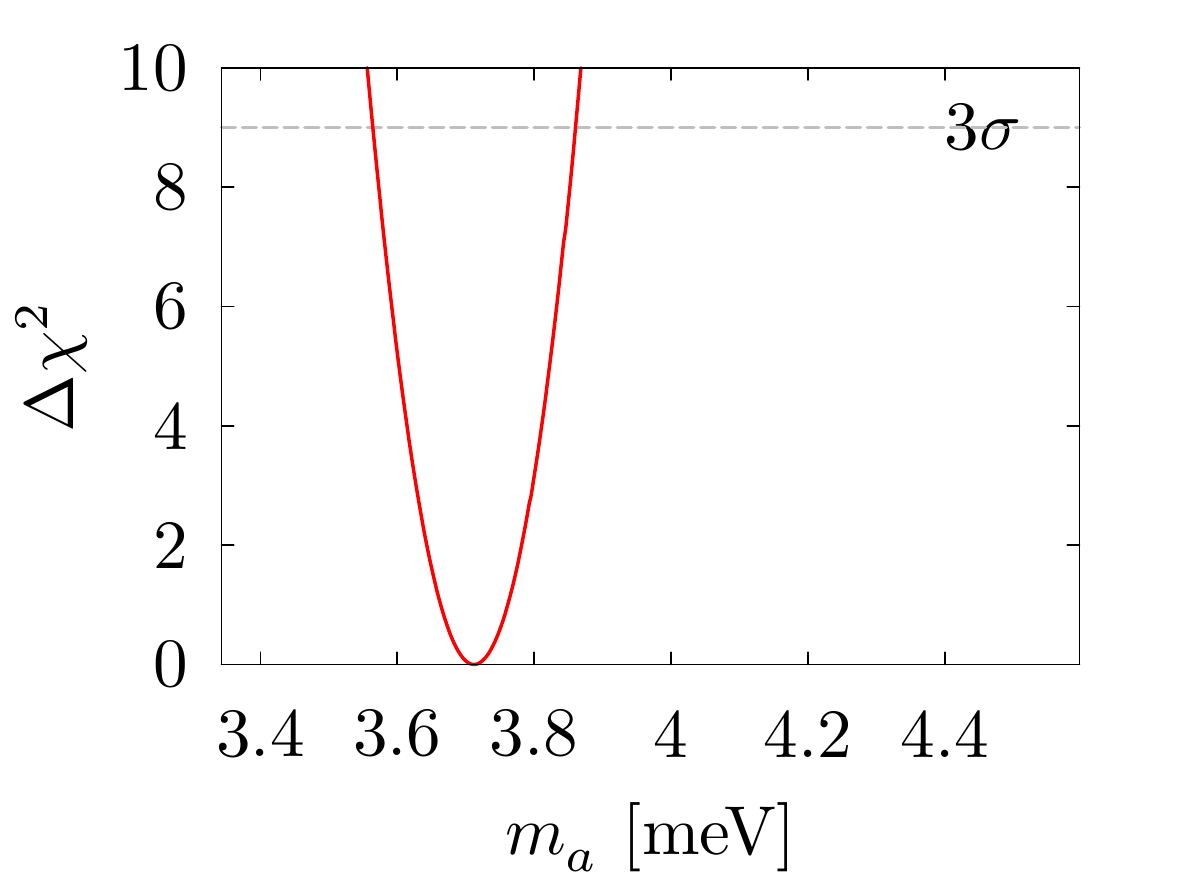}
\caption{The $\Delta\chi^2$ value against each model parameters for MOMENT. True values for the model parameters are used $(x,~\eta,~r,~m_a)=(-3.65,~1.13\pi,~0.511,~3.71~\text{meV})$. The range shown here is according to the $3\sigma$ uncertainty with NuFit4.0 results Eq.~(\ref{eq:3sigma_nufit4}):  $-5.475<x<-3.37$ (red band),~$0.455<\eta/\pi<1.545$ (dark grey band), ~$0.204<r<0.606$ (blue band),~$3.343<m_a/\text{meV}<4.597$ (yellow band).}%
\label{fig:model_1D}
\end{figure}

\begin{figure}[!h]
 \flushleft
\hspace{15mm}\includegraphics[width=0.32\textwidth]{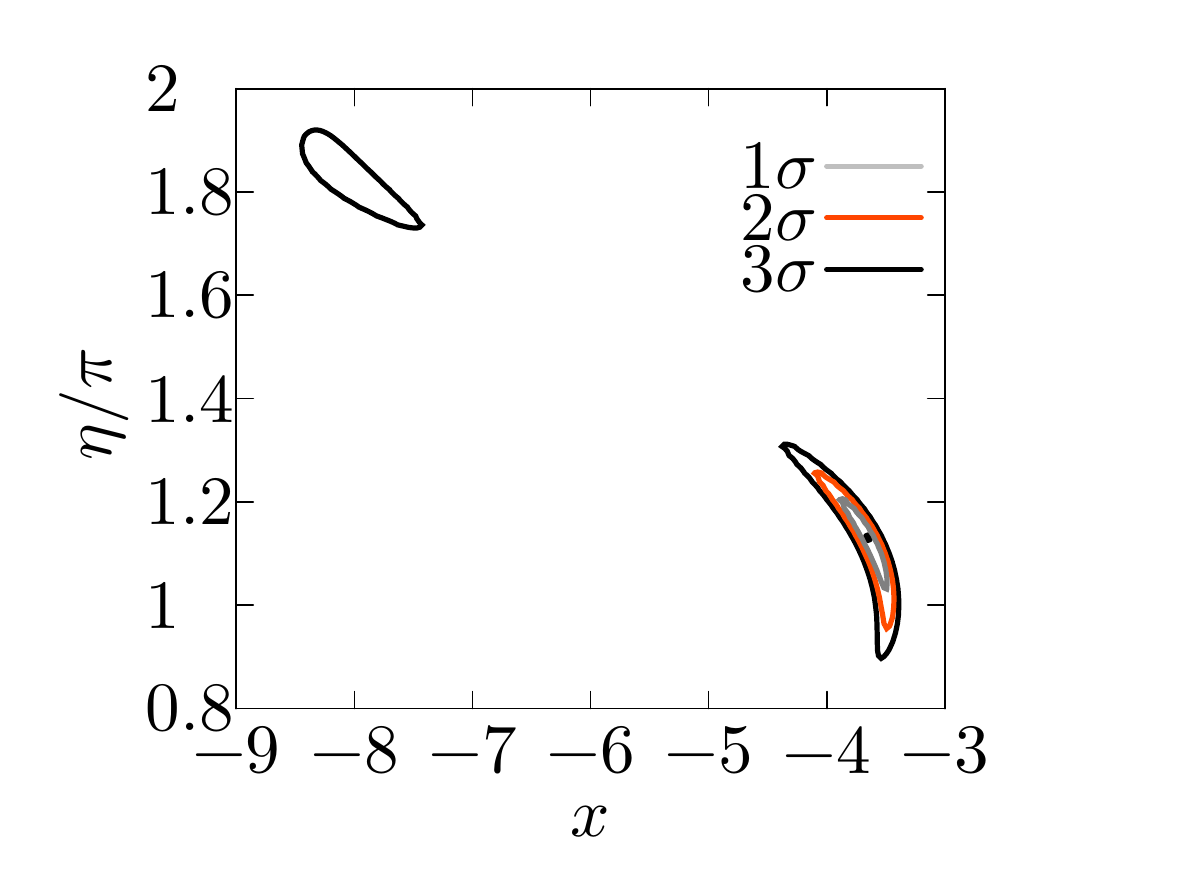}$~~~~~~$\\
\hspace{15mm} \includegraphics[width=0.32\textwidth]{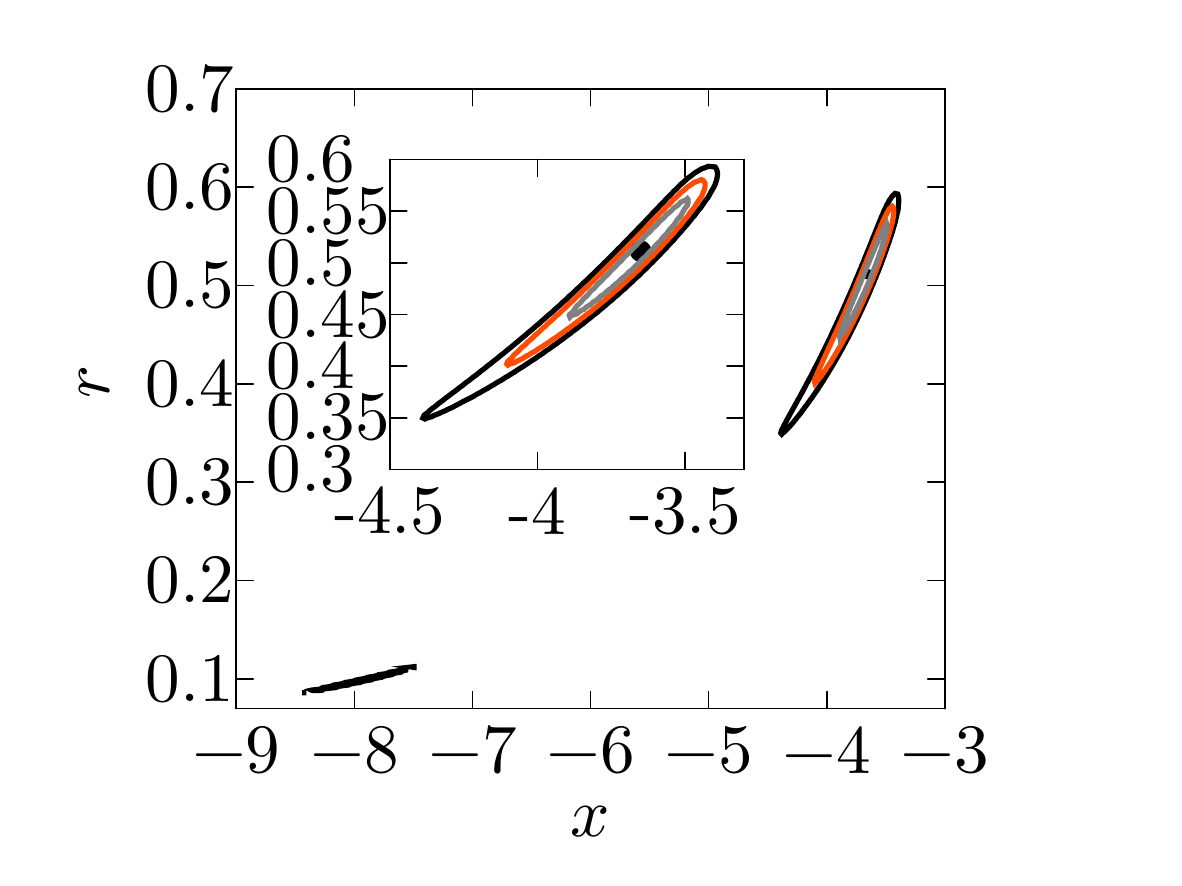}\hspace{-11mm}
 \includegraphics[width=0.32\textwidth]{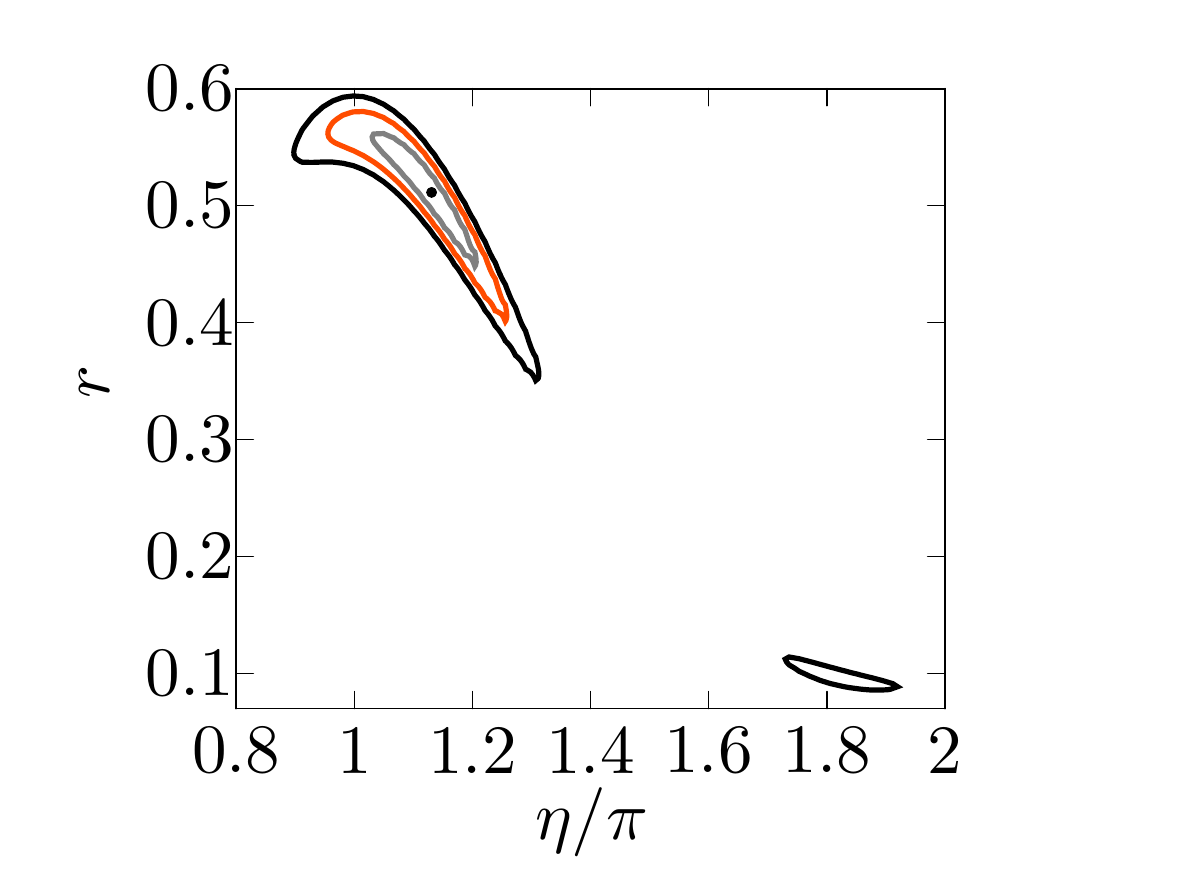}$~~~~~~$\\
\hspace{15mm} \includegraphics[width=0.32\textwidth]{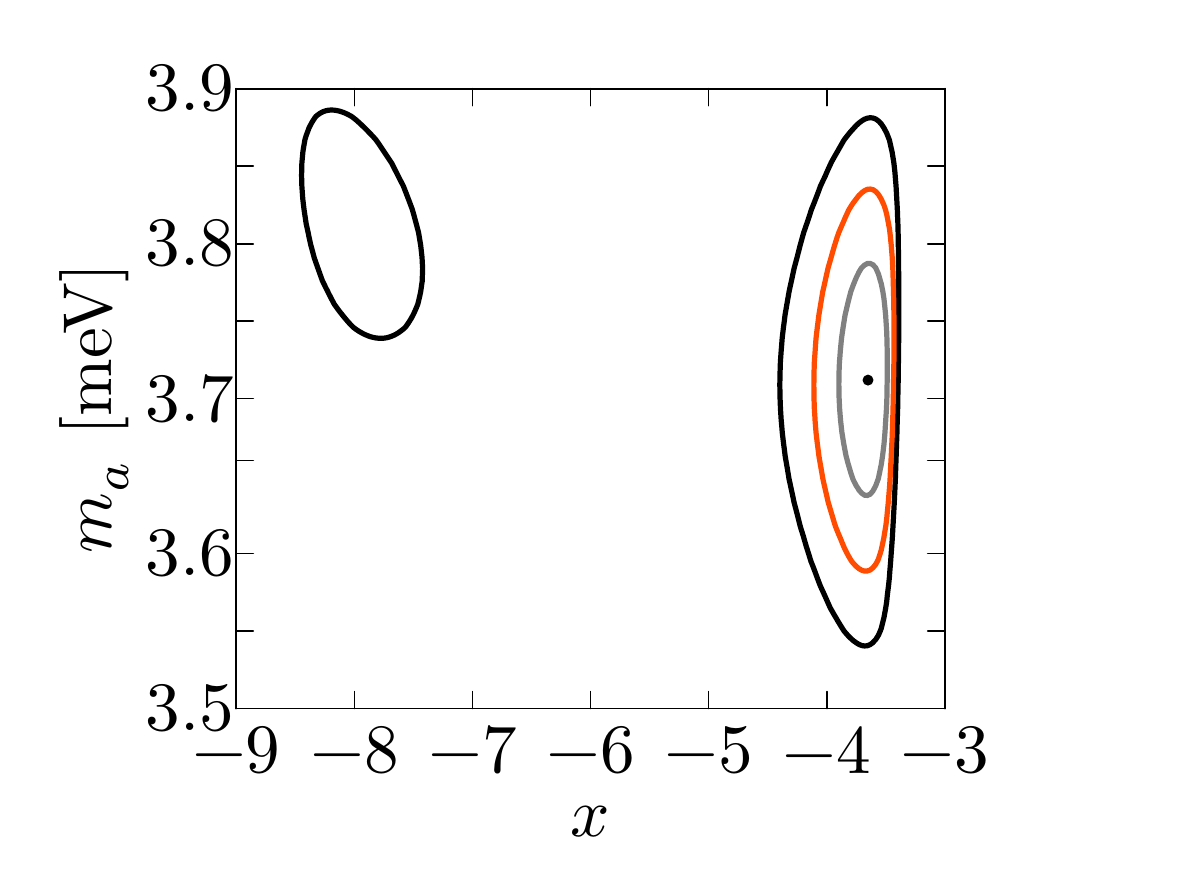}\hspace{-11mm}
 \includegraphics[width=0.32\textwidth]{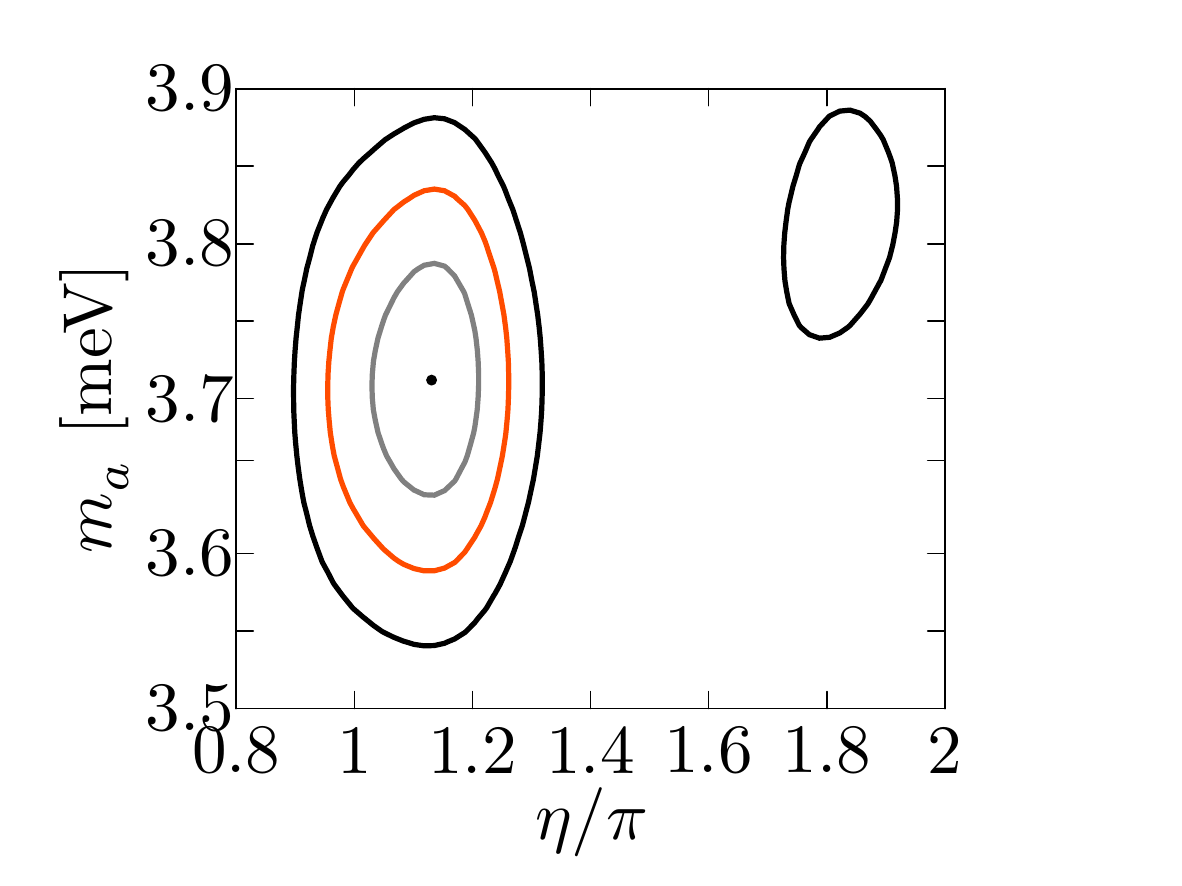}\hspace{-11mm}
 \includegraphics[width=0.32\textwidth]{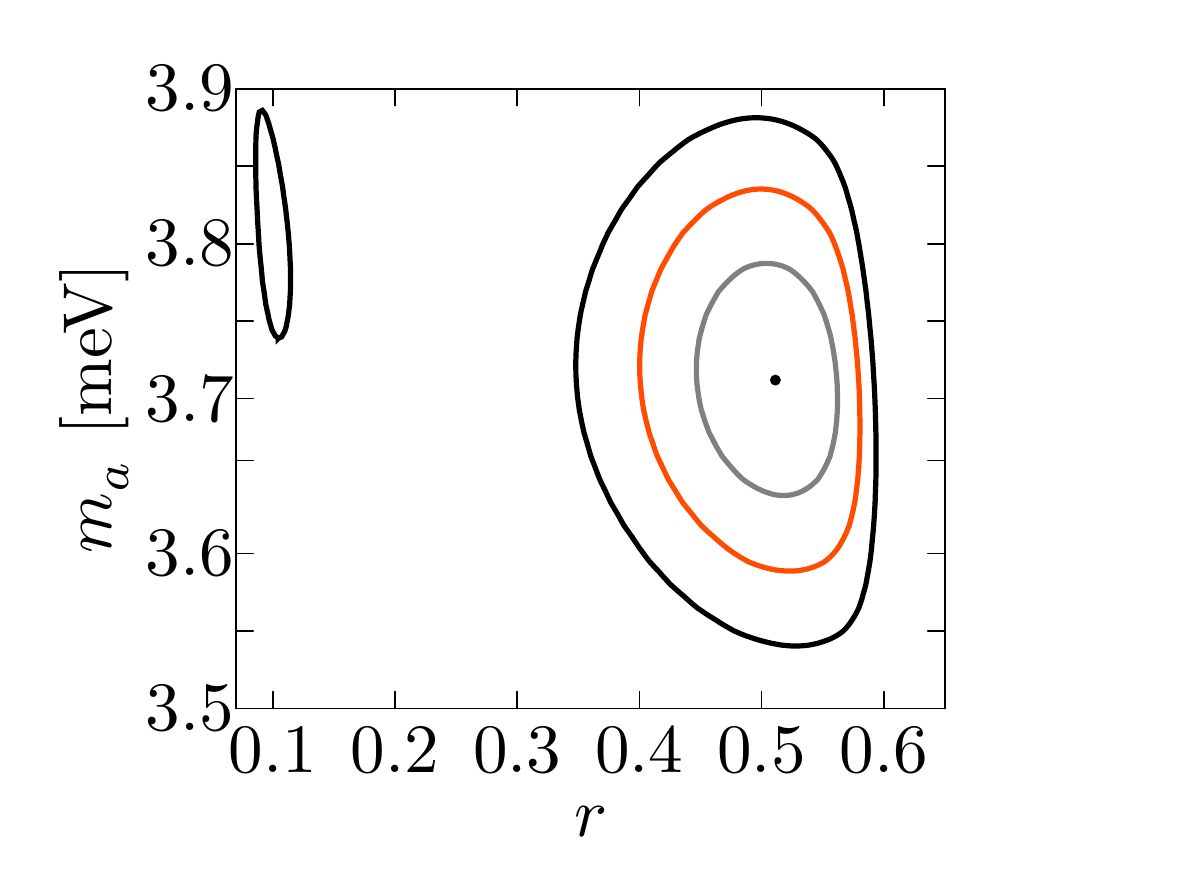}
 \caption{\label{fig:model_2D}Precision measurements of any two model parameters in the framework of three neutrino oscillations taking uncertainties of the current global fit results, for MOMENT, at $1\sigma$ (gray), $2\sigma$ (orange), $3\sigma$ (black) confidence level. True values for the model parameters are used $(x,~\eta,~r,~m_a)=(-3.65,~1.13\pi,~0.511,~3.71~\text{meV})$.}
\end{figure}

We study how model parameters can be constrained by MOMENT. For this purpose, we study the statistics quantity,
\begin{equation}\label{eq:MP_constrain}
%\Delta\chi^2=\sum_i\min_{\overrightarrow{\mathcal{M}}_{hyp.}}\chi^2(\mu_i(\overrightarrow{\mathcal{M}}_{hyp.}),n_i(\overrightarrow{\mathcal{M}}_{true}))-\sum_i\min_{\overrightarrow{\mathcal{M}}_{b.f.}}\chi^2(\mu_i(\overrightarrow{\mathcal{M}}_{b.f.}),n_i(\overrightarrow{\mathcal{M}}_{true})),
\Delta\chi^2=\sum_i\chi^2(\mu_i(\overrightarrow{\mathcal{M}}_{hyp.}),n_i(\overrightarrow{\mathcal{M}}_{true}))-\sum_i\chi^2(\mu_i(\overrightarrow{\mathcal{M}}_{b.f.}),n_i(\overrightarrow{\mathcal{M}}_{true})),
\end{equation}
where $\overrightarrow{\mathcal{M}}_{hyp.}$ is the hypothesis, $\overrightarrow{\mathcal{M}}_{true}$ is the true values, and $\overrightarrow{\mathcal{M}}_{b.f.}$ is the best fit. Here $\overrightarrow{\mathcal{M}}_{b.f.}$ is exactly $\overrightarrow{\mathcal{M}}_{true}$.
We show our result in Figs.~\ref{fig:model_1D} and \ref{fig:model_2D}. We set the true values at the $(x,~\eta,~r,~m_a)=(-7/2,~\pi,~0.553,~3.72~\text{meV})$, which is the best fit with NuFit4.0 results. And the range for each panel is the $3\sigma$ uncertainty with NuFit4.0 results Eq.~(\ref{eq:3sigma_nufit4}):  $-5.475<x<-3.37$ (red band),~$0.455<\eta/\pi<1.545$ (dark grey band), ~$0.204<r<0.606$ (blue band),~$3.343<m_a/\text{meV}<4.597$ (yellow band). At $3\sigma$ confidence level, the uncertainty of the model parameter $x$ lies roughly from $-4.25$ to $-3.5$. For the model parameter $\eta$, it ranges from $\sim0.925\pi$ to $\sim1.275\pi$. 
The $3\sigma$ errors for $r$ and $m_a$ are about $0.36<r<0.58$ and $3.5\text{meV}<m_a<3.85\text{meV}$. 
Compared to the result shown in Eq.~(\ref{eq:3sigma_nufit4}), we see the parameter with the least improvement is $r$, for which the $3\sigma$ uncertainty is improved by a factor of $2$. 
%We also compare the result shown in Ref.~\cite{Ding:2019zhn}, and find the performance of MOMENT is similar to that of DUNE ($-4.2<x<-3.5$, $0.92<\eta/\pi<1.28$, $0.4<r<0.6$ and $3.56<m_a/\text{meV}<3.86$ for $3\sigma$ uncertainties), but worse than that of T2HK ($-3.8<x<-3.5$, $0.92<\eta/\pi<1.18$, $0.45<r<0.6$ and $3.56<m_a/\text{meV}<3.86$ for $3\sigma$ uncertainties).

In Fig.~\ref{fig:model_2D}, we show $1\sigma$ (gray), $2\sigma$ (light-orange) and $3\sigma$ (black) contours on the plane spanned by any two of model parameters. We see a strong correlation among $x$, $\eta$ and $r$, which is consistent with Eq.~(\ref{eq:mnu}). In Eq.~(\ref{eq:mnu}), we see these three parameters joint in the matrix for the neutrino solar mass. As a result these degeneracies can be resolved by precision measurement of solar mixing angle $\theta_{12}$ or solar mass-square splitting $\Delta m_{21}^2$. This degeneracy problem has also addressed by simulation results in other LBL experimental configurations, and is known to be resolved by the precision measurement of $\theta_{12}$~\cite{Ding:2019zhn}.

%%%%%%%%%%%%%%%%%%%%%%%%%%%%%%%%%%%%%%%%%%%%%%%%%%%%%%%
\subsection{Projection on the standard-parameter space}\label{sec:projection}

\begin{figure}[!h]%
\centering
\includegraphics[width=2.7in]{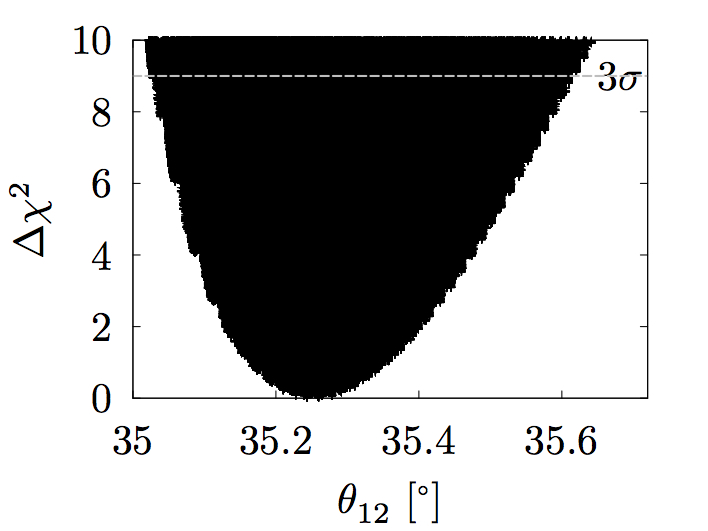}
\includegraphics[width=2.7in]{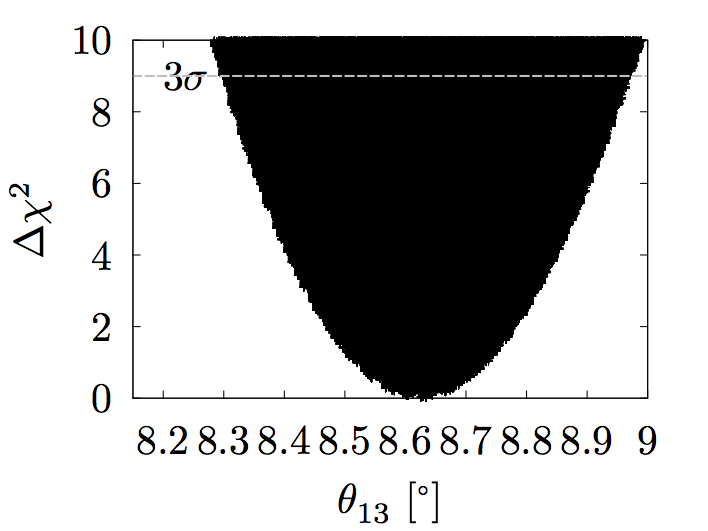}\\
\includegraphics[width=2.7in]{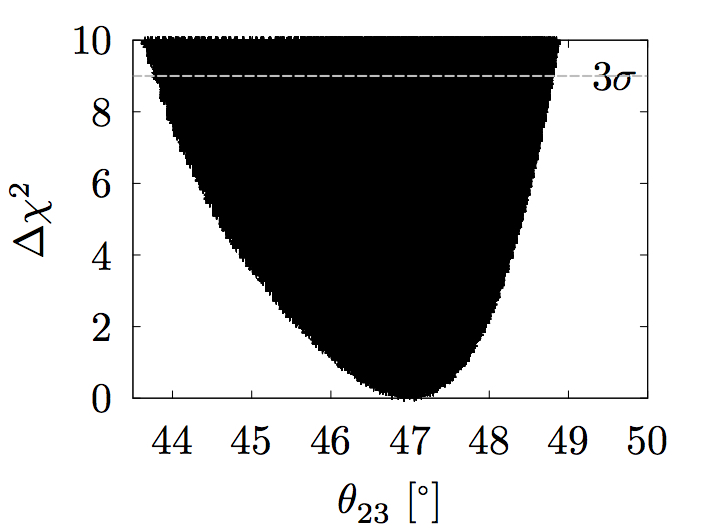}
\includegraphics[width=2.7in]{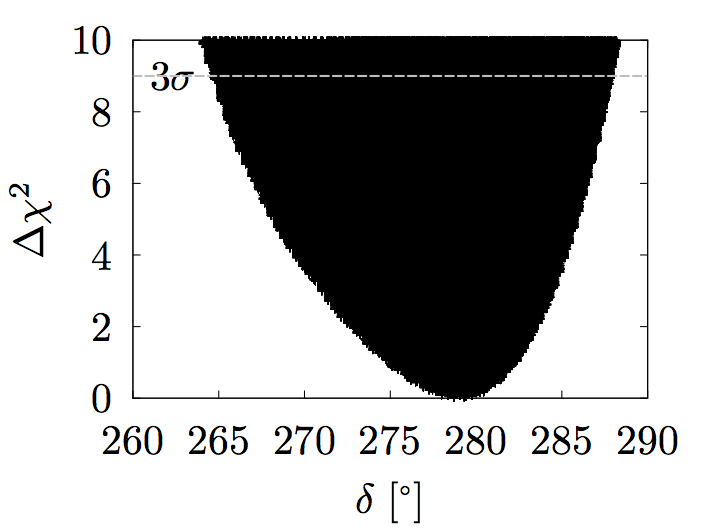}\\
\includegraphics[width=2.7in]{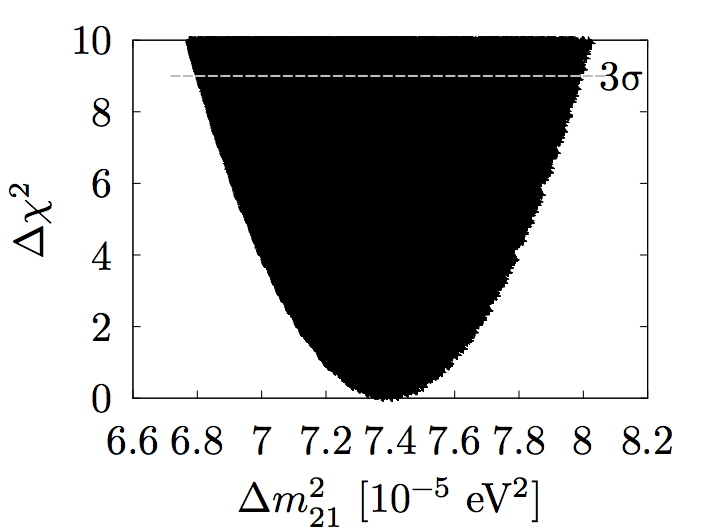}
\includegraphics[width=2.7in]{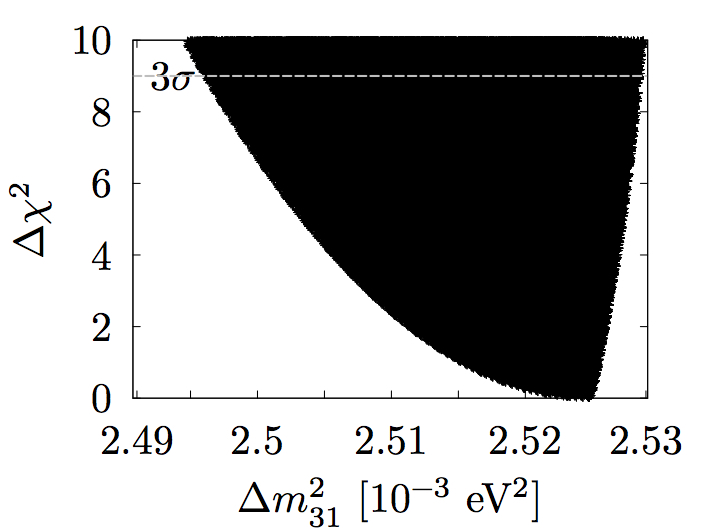}
\caption{The $\Delta\chi^2$ value against $\theta_{12}$ (upper left), $\theta_{13}$ (upper right), $\theta_{23}$ (middle left), $\delta$ (middle right), $\Delta m_{21}^2$ (lower left) and $\Delta m_{31}^2$ (lower right), for MOMENT experiment, assuming the tri-direct model.}%
\label{fig:PMNS_1D}
\end{figure}

In Fig.~\ref{fig:PMNS_1D}, we project points inside the  $3\sigma$ sphere from the 4-dimension model-parameter space on each oscillation parameters with their $\Delta\chi^2$ values (y-axis). Though MOMENT is not sensitive to $\theta_{12}$, we see that this parameter is well constrained to be better than that of NuFit4.0 result. The uncertainty for $\theta_{13}$ and $\Delta m_{21}^2$ are almost the same as the $3\sigma$ errors NuFit4.0. %
The asymmetry for $\theta_{12}$, $\theta_{23}$ and $\Delta m_{31}^2$ is passed by the same feature of $x$, $\eta$, and $m_a$.

\begin{figure}[!h]%
\centering
\includegraphics[width=2.7in]{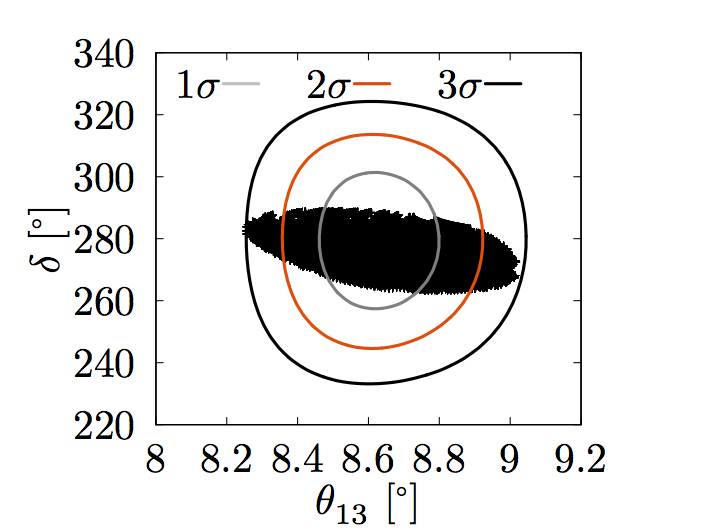}
\includegraphics[width=2.7in]{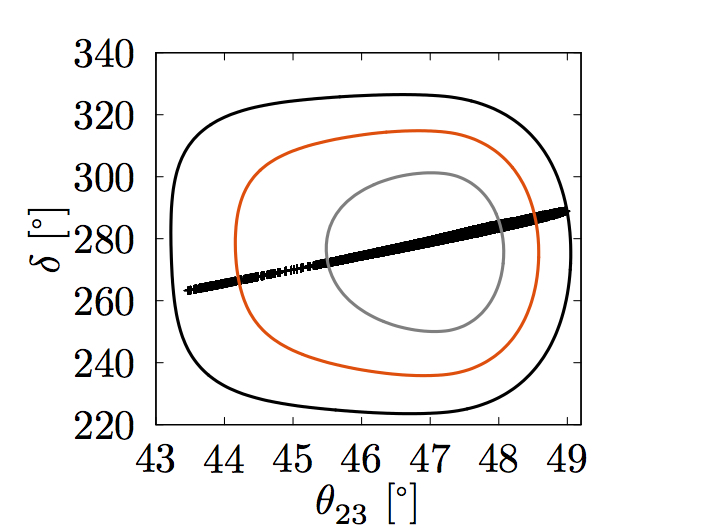}\\
\includegraphics[width=2.7in]{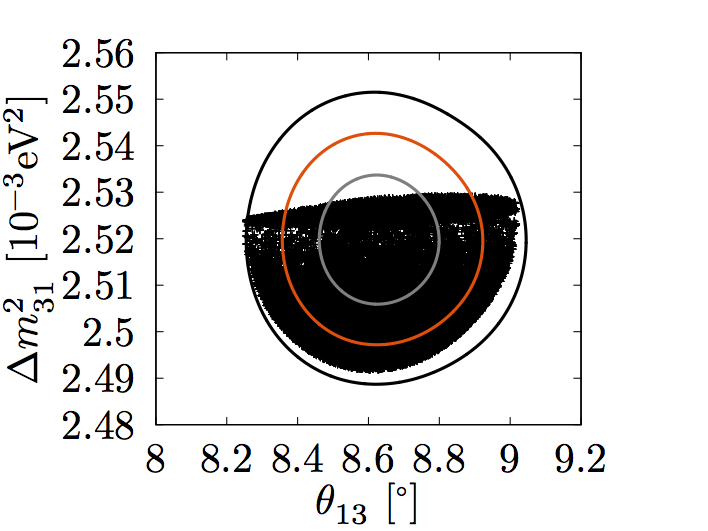}
\includegraphics[width=2.7in]{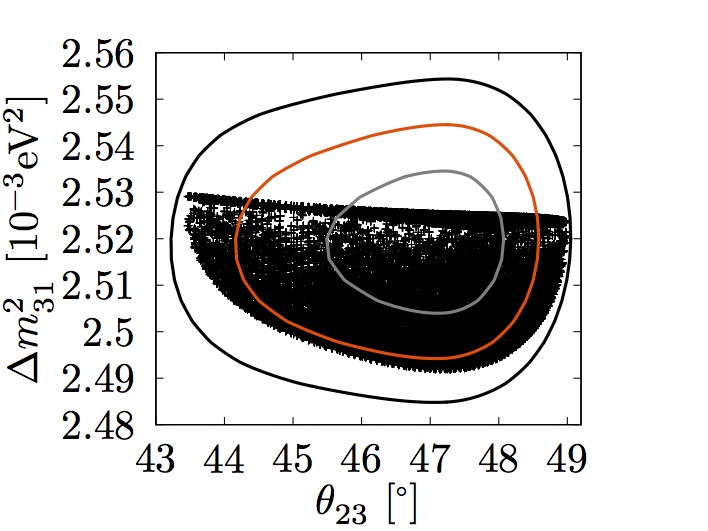}
\caption{The points at the 4-dimension sphere at the $3\sigma$ projected on $\theta_{13}$-$\delta$ (upper-left), $\theta_{23}$-$\delta$(upper-right), $\theta_{13}$-$\Delta m^2_{31}$ (lower-left), $\theta_{13}$-$\Delta m_{31}^2$(lower-right) for MOMENT experiment. We also present the $1\sigma$ (grey), $2\sigma$ (orange), and $3\sigma$ (black) contours without the restriction of TDLS.}%
\label{fig:PMNS_2D}
\end{figure}

In Fig.~\ref{fig:PMNS_2D}, we project the $3\sigma$ sphere from the 4-dimension model-parameter space to the two-dimension plane spanned by the standard oscillation parameters. We see that under the TDLS model, $\delta$ and $\Delta m_{31}^2$ are constrained better than those without assuming TDLS models by about a factor of $2$. The uncertainty for $\theta_{23}$ is slightly better when TDLS is assumed. The $3\sigma$ uncertainty for $\theta_{13}$ is roughly the same between with and without assuming TDLS models. %{\color{red}[checking]This is caused by the tension between the simulated data for MOMENT and global fit result as priors. As a result, the uncertainty for $\theta_{23}$ is twisted, once the $3\sigma$ bounds for the model parameters predicted by the prior for $\Delta m_{21}^2$ are inconsistent to those by the $\theta_{23}$ measurement of MOMENT, and if the constraint by the prior is much tighter than that by simulation data.}
The band feature in the $\theta_{23}$-$\delta$ panel can be understood by the expansions of $\cos\delta$ and $\sin\delta$ in Table~\ref{tab:parameters}: 
\begin{equation}\label{eq:cosdelta}
\cos\delta=\frac{ \cot 2 \theta_{23} \left[3x^2-\left(4x^2+ x+1\right)\cos^2\theta_{13}\right]}{\sqrt{3} \left|x\right| \sin \theta_{13} \sqrt{\left(5x^2+2x+2\right)\cos^2\theta_{13}-3x^2}}\,,
\end{equation}
and
\begin{equation}\label{eq:sindelta}
\sin\delta= \pm\csc 2 \theta_{23} \sqrt{1+\frac{\left(x^2+x+1\right)^2 \cot ^2\theta_{13} \cos ^22 \theta_{23}}{3x^2 \left[3x^2 \tan 2\theta_{13}-2 \left(x^2+x+1\right)\right]}}~\,,
\end{equation}
with ``$+$'' for $x\cos\psi>0$ and ``$-$'' for $x\cos\psi<0$.

Considering $\theta_{23}\sim45^\circ$, we have
\begin{equation}
\begin{array}{c}
\cos\delta\propto\cot 2\theta_{23}=\frac{\cos 2\theta_{23}}{\sin 2\theta_{23}}, \\ \sin\delta \propto \pm \csc 2\theta_{23} =\pm\frac{1}{\sin 2\theta_{23}}.
\end{array}
\end{equation}
Therefore, we have 
\begin{equation}\label{eq:delta_th23}
\tan\delta\propto1/\cos2\theta_{23}.
\end{equation} 
Eq.~(\ref{eq:delta_th23}) predicts that if $\theta_{23}=45^\circ$, $\delta=90^\circ$ or $270^\circ$, which is also confirmed in the $\theta_{23}$-$\delta$ panel of Fig.~\ref{fig:PMNS_2D}. On the other hand, due to the poor sensitivity to the solar angle of MOMENT, we do not see the result reflecting the sum rule Eq.~(\ref{eq:correlation_mix_angles}).

%{\color{red}
%\begin{figure}[!h]%
%\flushleft
%\hspace{38mm}\includegraphics[width=2.in]{12_13.jpg}\\
%\hspace{38mm}\includegraphics[width=2.in]{x_12.jpg}\hspace{-11mm}
%\includegraphics[width=2.in]{x_13.jpg}
%\caption{The points at the 4-dimension sphere at the $3\sigma$ projected on $\cos^2\theta_{12}$-$\cos^2\theta_{13}$ (upper), $%\cos^2\theta_{12}$-$x$ (lower-left) and $\cos^2\theta_{13}$-$x$ (lower-right) for MOMENT experiment. }%
%\label{fig:12_13}
%\end{figure}

%\begin{figure}[!h]%
%\centering
%\includegraphics[width=3.in]{12_13.jpg}\\
%\caption{The points at the 4-dimension sphere at the $3\sigma$ projected on $\cos^2\theta_{12}$-$\cos^2\theta_{13}$ for MOMENT experiment. }%
%\label{fig:12_13}
%\end{figure}

%We also investigate the sum rule Eq.~\ref{eq:correlation_mix_angles} with simulated MOMENT data in Fig.~\ref{fig:12_13}. We project the points at the $4$-dimension sphere at the $3\sigma$ on the $\cos^2\theta_{12}$-$\cos^2\theta_{13}$ plane. We see the tendency is roughly along with $\cos^2\theta_{12}=\cos^2\theta_{13}$, and consistent with the sum rule Eq.~\ref{eq:correlation_mix_angles}.The smaller region around $(cos^2\theta_{12},cos^2\theta_{13})=(0.3,~0.5)$ is because of the degeneracy we see in Fig.~\ref{fig:PMNS_2D}.}

\section{Conclusion}
\label{sec:conclusion}

We have studied how we can extend our knowledge on the flavor symmetry with MOMENT, using eight channels of neutrino oscillations ($\nu_e\rightarrow \nu_e$, $\nu_e\rightarrow \nu_{\mu}$, $\nu_{\mu} \rightarrow \nu_e$, $\nu_{\mu} \rightarrow \nu_{\mu}$ and their CP-conjugate partners) with the help of the following detection processes in a Gd-doped water Cherenkov detector: $\nu_e + n \rightarrow p + e^-$, $\bar{\nu}_{\mu} + p \rightarrow n + \mu^+$, $\bar{\nu}_e + p \rightarrow n + e^+$, and $\nu_{\mu} + n \rightarrow p + \mu^- $. We have analyzed the physics potential of MOMENT on littlest seesaw models in the tri-direct approach given in Eq.~(\ref{eq:mnu}) as a case study. 
%
%Current proposed littlest seesaw models in tri-direct approach (TDLS) Eq.~(\ref{eq:mnu}) well describes the current global-fit result: NuFit4.0 (shown in Tab.~\ref{tab:nufit4.0}). %The current understanding of this model has been summarised in Tab.~\ref{tab:model_fit}: except for $\theta_{23}$ and $\delta$, TDLS well describes the current oscillation parameters. 

We have studied the exclusion ability to TDLS models for MOMENT. We found that $\theta_{23}$ and $\delta$ are the most important parameters to exclude this model, though some contributions from $\theta_{13}$ and $\Delta m_{31}^2$ are also seen. We noticed that the precision measurement in MOMENT of $\theta_{23}$ and $\delta$ can exclude this model with more than $5\sigma$ significance, if the best fit of NuFit4.0 is confirmed. We also presented the constraint on model parameters with simulated MOMENT data. We have found MOMENT data can improve the $3\sigma$ uncertainty by at least a factor of two, compared to those by NuFit4.0 results shown in Eq.~(\ref{eq:3sigma_nufit4}). We have found the degeneracy problem, which is caused by the poor measurement of $\theta_{12}$. This degeneracy problem has been addressed in Ref.~\cite{Ding:2019zhn}. We projected the $3\sigma$ sphere from the model-parameter space to the oscillation-parameter space. 
Finally, we have found that the sum rule between $\theta_{23}$ and $\delta$: $\tan\delta\propto1/\cos2\theta_{23}$ (for $\theta_{23}\sim 45^\circ$) predicted by Eqs.~(\ref{eq:cosdelta}) and (\ref{eq:sindelta}) can be checked by MOMENT.

Finally, we come to the conclusion that $\theta_{23}$ and $\delta$ are the most important  parameters in the standard neutrino mixing framework to understand the underlying TDLS model. It is not only because they are the only two parameters, of which the model prediction deviates from the best fit of NuFit4.0 by more than $1\sigma$, but also because they can exclude this model at the $5\sigma$ confidence level as soon as the best fit values are confirmed in the future global analysis. As a result, to optimize the experimental design at MOMENT for the purpose of understanding the TDLS model, we need to aim at precision measurements of $\theta_{23}$  and $\delta$.

\section{Acknowledgement}%T.G.
This work is supported in part by the National Natural Science Foundation of China under Grant No. 11505301 and No. 11881240247. We appreciate Gui-Jun Ding's great help in understanding the tri-direct symmetry models. We would like to thank the accelerator working group of MOMENT for useful discussions and for kindly providing flux files for the MOMENT experiment. We finally acknowledge Dr.~Sampsa Vihonen's help to improve the readability of this paper. 

%\appendix 

% \bibliographystyle{unsrt}
\bibliography{tri-direct.bib}

\end{document}